%% file: main.tex
\documentclass[10pt,twocolumn,letterpaper]{article}

\usepackage{iccv}
\usepackage{times}
\usepackage{graphicx}
\usepackage{amsmath}
\usepackage{amssymb}
\usepackage{subfig}
\usepackage[bookmarks=false]{hyperref}
\usepackage[belowskip=-10pt,aboveskip=0pt]{caption} 
\usepackage{placeins} 

\newcommand{\mbf}[1]{\mathbf{#1}}
\newcommand{\mc}[1]{\mathcal{#1}}
\newcommand{\mb}[1]{\mathbb{#1}}

\newcommand{\z}{\mathbf{z}}
\newcommand{\x}{\mathbf{x}}

\newcommand{\q}{\mathbf{q}}

\newcommand{\pder}[2]{\frac{\partial#1}{\partial#2}}

\newcommand{\KL}{\operatorname{KL}}
\newcommand{\CE}{\operatorname{CE}}

\iccvfinalcopy 

\ificcvfinal\fi

\title{Video Compression With Rate-Distortion Autoencoders}

\author{Amirhossein Habibian, Ties van Rozendaal, Jakub M. Tomczak, Taco S. Cohen\\
Qualcomm AI Research\thanks{Qualcomm AI Research is an initiative of Qualcomm Technologies, Inc.}, Amsterdam, the Netherlands \\
{\tt\small \{habibian, ties, jtomczak, tacos\}@qti.qualcomm.com}
}

\begin{document}

\maketitle

\begin{abstract}
In this paper we present a deep generative model for lossy video compression.
We employ a model that consists of a 3D autoencoder with a discrete latent space and an autoregressive prior used for entropy coding.
Both autoencoder and prior are trained jointly to minimize a rate-distortion loss, which is closely related to the ELBO used in variational autoencoders.
Despite its simplicity, we find that our method outperforms the state-of-the-art learned video compression networks based on motion compensation or interpolation.
We systematically evaluate various design choices, such as the use of frame-based or spatio-temporal autoencoders, and the type of autoregressive prior.

In addition, we present three extensions of the basic method that demonstrate the benefits over classical approaches to compression.
First, we introduce \emph{semantic compression}, where the model is trained to allocate more bits to objects of interest.
Second, we study \emph{adaptive compression}, where the model is adapted to a domain with limited variability, \eg videos taken from an autonomous car, to achieve superior compression on that domain.
Finally, we introduce \emph{multimodal compression}, where we demonstrate the effectiveness of our model in joint compression of multiple modalities captured by non-standard imaging sensors, such as quad cameras. We believe that this opens up novel video compression applications, which have not been feasible with classical codecs.
\end{abstract}

\section{Introduction}
\label{sec:intro}
\input{intro}



\section{Related Work}
\label{sec:related_work}

\input{related-work}

\section{Rate-Distortion Autoencoders \& VAEs}
\label{sec:theory}
\input{theory}

\section{Methodology}
\label{sec:methods}
\input{methods}

\section{Experiments}
\label{sec:experiments}
\input{experiments}


\section{Conclusion}
\label{sec:conclusion}
\input{conclusion}


{\small

\small}

\clearpage
\section{Supplementary Material}
\input{supplementary}

\end{document}

%% file: intro.tex
In recent years, tremendous progress has been made in generative modelling.
Although much of this work has been motivated by potential future applications such as model based reinforcement learning, \emph{data compression} is a very natural application that has received comparatively little attention.
Deep learning-based video compression in particular has only recently started to be explored \cite{chenLearningVideoCompression2018, rippelLearnedVideoCompression2018, wuVideoCompressionImage2018}.
This is remarkable because improved video compression would have a large economic impact: it is estimated that very soon, $80\%$ of internet traffic will be in the form of video \cite{ciscoZettabyteEraTrends2017}.

\begin{figure}[t!]
	\centering
    \includegraphics[width=0.95\linewidth]{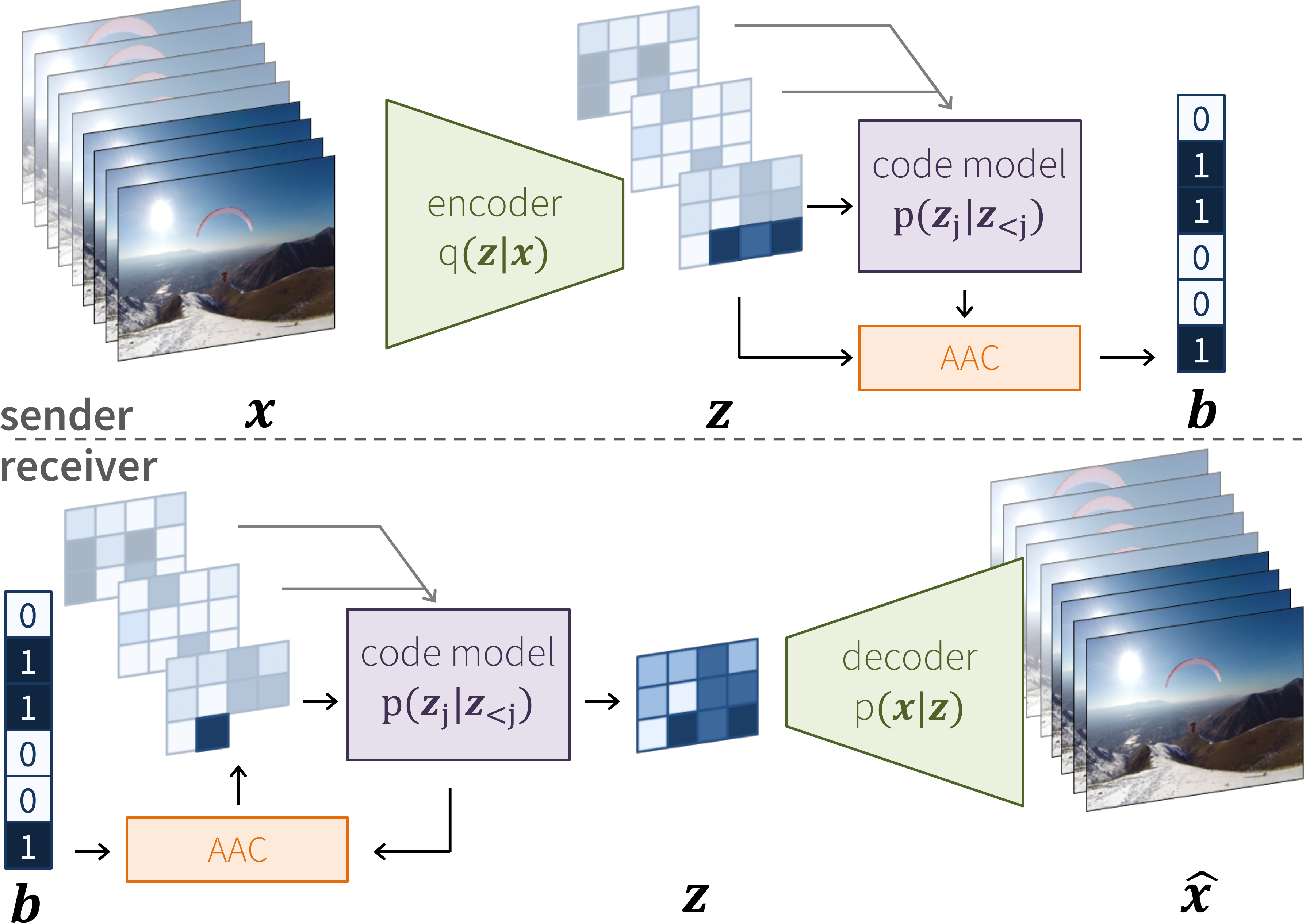}
    \caption{Overview of the proposed compression inference pipeline. The encoder encodes a sequence of frames $\mathbf{x}$ into a sequence of quantized latent variables $\mathbf{z}$. A code model $p(\mathbf{z}_{t}|\mathbf{z}_{<t})$ is used to transform $\mathbf{z}$ into a bitstream $\mathbf{b}$ using adaptive arithmetic coding (AAC). On the receiver side, the bitstream is used to reconstruct $\mathbf{z}$ which is then lossily decoded into $\mathbf{\hat{x}}$.
    }
    \label{fig:model-inference}
\end{figure}

In this paper, we present a simple yet effective and theoretically grounded method for video compression that can serve as the basis for future work in this nascent area.
Our model consists of off-the-shelf components from the deep generative modelling literature, namely autoencoders (AE) and autoregressive models (ARM). 
Despite its simplicity, the model outperforms all methods to which a direct comparison is possible, including substantially more complicated approaches.

On the theoretical side, we show that our method, as well as state-of-the-art image compression methods \cite{mentzerConditionalProbabilityModels2018} can be interpreted as VAEs \cite{kingmaAutoEncodingVariationalBayes2013, rezendeStochasticBackpropagationApproximate2014} with a discrete latent space and a deterministic encoder.
The VAE framework is an especially good fit for the problem of \emph{lossy} compression, because it provides a natural mechanism for trading off \emph{rate} and \emph{distortion}, as measured by the two VAE loss terms \cite{alemiFixingBrokenELBO2017}.
However, as we will argue in this paper, it is not beneficial for the purpose of compression to use a stochastic encoder (approximate posterior) as usually done in VAEs, because any noise added to the encodings results in increased bitrate without resulting in an improvement in distortion \cite{hintonKeepingNeuralNetworks}.

On the experimental side, we perform an extensive evaluation of several architectural choices, such as the use of 2D or 3D autencoders, and the type of autoregressive prior.
Our best model uses a ResNet \cite{heDeepResidualLearning2016} autoencoder with 3D convolutions, and a temporally-conditioned gated PixelCNN~\cite{vandenoordConditionalImageGeneration2016a} as prior.
We benchmark our method against existing learned video compression methods, and show that it achieves better rate/distortion.
We also find that our method outperforms the state-of-the-art traditional codecs when these are used with restricted settings, as it is done in previous work, but more work remains to be done before it can be claimed that these learned video compression methods suppress traditional codecs under optimal settings.

Additionally, we introduce several extensions of our method that highlight the benefits of using learned video codecs.
In \emph{semantic compression}, we bridge the gap between semantic video understanding and compression by learning to allocate more bits to objects from categories of interest,~\ie, people. During training, we weight the rate and distortion losses to ensure a high quality reconstruction for regions of interest extracted by off-the-shelf object detection or segmentation networks, such as Mask R-CNN\cite{heMaskRCNN2017}.

We also demonstrate \emph{adaptive compression}, where the model is trained on a specific domain, either before or after deployment, to fine-tune it to the distribution of videos it is actually used for.
We show that adaptive compression of footage from autonomous cars can result in large improvement in terms of rate and distortion.
With classical codecs, finetuning for a given domain is often not feasible.

Finally, we show that our method is very effective in joint compression of multiple modalities, which exist in videos from depth, stereo, or multi view cameras. By utilizing the siginifcant redundancy, which exist in multimodal videos, our model outperforms HEVC/H.265 and AVC/H.264 by a factor of 4.

The main contributions of this paper are:
\emph{i)} We present a simple yet effective and theoretically grounded method for video compression that can serve as the basis for future work.
\emph{ii)} We clarify theoretically the relation between rate-distortion autoencoders and VAEs.
\emph{iii)} We introduce semantic compression to bridge the gap between semantic video understanding and compression.
\emph{iv)} We introduce adaptive compression to adapt a compression model to the domain of interest.
\emph{v)} We introduce multimodal compression to jointly compress multiple modalities, which exist in a video using a deep video compression network.

The rest of the paper is organized as follows.
In the next section, we discuss related work on learned image and video compression.
Then, in section \ref{sec:theory}, we discuss the theoretical framework of learned compression using rate-distortion autoencoders, as well as the relation to variational autoencoders.
In section \ref{sec:methods} we discuss our methodology in detail, including data preprocessing and autoencoder and prior architecture.
We present experimental results in section \ref{sec:experiments}, comparing our method to classical and learned video codecs, evaluating semantic compression, adaptive compression, and multimodal compression.
Section \ref{sec:conclusion} concludes the paper.

%% file: related-work.tex
\textbf{Learned Image Compression} 
Deep neural networks are the state-of-the-art in image compression outperforming all traditional compression algorithms such as BPG and JPEG2000.
They often embed an input image into a low dimensional representation using fully convolutional~\cite{mentzerConditionalProbabilityModels2018} or recurrent networks~\cite{baigLearningInpaintImage2017, johnstonImprovedLossyImage2017, todericiFullResolutionImage2017}.
The image representation is quantized by soft scalar quantization~\cite{agustssonSofttoHardVectorQuantization2017}, stochastic binarization~\cite{todericiFullResolutionImage2017}, or by adding uniform noise~\cite{balleEndtoendOptimizedImage2016a} to approximate the non-differentiable quantization operation.
The discrete image representation can be further compressed by minimizing the entropy during~\cite{chenVariationalLossyAutoencoder2016, mentzerConditionalProbabilityModels2018} or after training~\cite{balleEndtoendOptimizedImage2016a, balleVariationalImageCompression2018, liLearningConvolutionalNetworks2018}.
The models are typically trained to minimize the mean squared error between original and decompressed images or by using more perceptual metrics such as MS-SSIM~\cite{rippelRealTimeAdaptiveImage2017a} or adversarial loss~\cite{santurkar2018generative}.

The closest to us is the rate-distortion autoencoder proposed in~\cite{mentzerConditionalProbabilityModels2018} for image compression. We extend this work to video compression by: \emph{i)} proposing a gated conditional autoregressive prior using 2D convolutions \cite{vandenoordConditionalImageGeneration2016a} with, optionally, a recurrent neural net for better entropy estimation over time, \emph{ii)} encoding/decoding multiple frames by using 3D convolutions, \emph{iii}) simplifying the model and training by removing the spatial importance map~\cite{liLearningConvolutionalNetworks2018} and disjoint entropy estimation, without any loss on compression performance.

\textbf{Learned Video Compression}
Video compression shares many similarities with image compression, but the large size of video data, and the very high degree of redundancy create new challenges \cite{hanDeepProbabilisticVideo2018, pessoaEndtoEndLearningVideo2018, rippelLearnedVideoCompression2018, wuVideoCompressionImage2018}.
One of the first deep learning-based approaches proposes to model video autoregressively with a RNN-conditioned PixelCNN \cite{kalchbrennerVideoPixelNetworks2016}.
While being powerful and flexible, this model scales rather poorly to larger videos, and can only be used for lossless compression.
Hence, we employ this method for lossless compression of latent codes, which are much smaller than the video itself.
An extension of this method was proposed in \cite{chenLearningVideoCompression2018} where blocks of pixels are modeled in an autoregressive fashion and the latent space is binarized like in \cite{todericiFullResolutionImage2017}.
The applicability of this approach is rather limited since it is still not very scalable, and introduces artifacts in the boundary between blocks, especially for low bit rates.

The method described in \cite{wuVideoCompressionImage2018} compresses videos by first encoding key frames, and then interpolating them in a hierarchical manner.
The results are on par with AVC/H.264 when inter-frame compression is limited to only few (up to $12$) frames.
However, this method requires additional components to handle a context of the predicted frame. 
In our approach, we aim at learning these interactions through 3D convolutions instead.
In \cite{hanDeepProbabilisticVideo2018} a stochastic variational compression method for video was presented.
The model contains a separate latent variable for each frame, and for the inter-frame dependencies, and uses the prior proposed in \cite{balleVariationalImageCompression2018}.
By contrast, we use a simpler model with a single latent space, and use a deterministic instead of stochastic encoder.

Very recently the video compression problem was attacked by considering flow compression and residual compression \cite{luDVCEndtoendDeep2018, rippelLearnedVideoCompression2018}.
The additional components for flow and residual modeling allow to improve distortion in general, however, for low bit rates the proposed method is still outperformed by HEVC/H.265 on benchmark datasets.
Nevertheless, we believe that these ideas are promising and may be able to further improve the result presented in this paper.

%% file: theory.tex

Our general approach to lossy compression is to learn a latent variable model in which the latent variables capture the important information that is to be transmitted, and from which the original input can be approximately reconstructed.
We begin by defining a joint model of data $\x$ and \emph{discrete} latent variables $\z$,
\begin{equation}
    p_\theta(\mbf{x}, \mbf{z}) = p_\theta(\mbf{z}) p_\theta(\mbf{x} | \mbf{z})
\end{equation}
In the next section we will discuss the specific form of $p_\theta(\mbf{z})$ (the prior / code model) and $p_\theta(\mbf{x} | \mbf{z})$ (the likelihood / decoder), both of which will be defined in terms of deep networks, but for now we will consider them as general parameterized distributions.

Since the likelihood $\log p_\theta(\x) = \log \int p_\theta(\z) p_\theta(\x | \z) d\z$ is intractable, one optimizes the variational bound \cite{bishopPatternRecognitionMachine2006, wainwrightGraphicalModelsExponential2007a},
\begin{equation}
    \label{eq:bound}
    -\log p(\mbf{x}) \leq E_{q}[-\log p(\mbf{x} | \mbf{z})] + \operatorname{KL}[q(\mbf{z} | \mbf{x}) | p(\mbf{z})],
\end{equation}
where $q(\z|\x)$ is a newly introduced approximate posterior.
In the VAE \cite{kingmaAutoEncodingVariationalBayes2013,rezendeStochasticBackpropagationApproximate2014}, one uses neural networks to parameterize both $q(\z | \x)$ and $p(\x|\z)$, which can thus be thought of as the encoder and decoder part of an autoencoder.

The VAE is commonly interpreted as a regularized auto-encoder, where the first term of the loss measures the reconstruction error and the KL term acts as a regularizer \cite{kingmaAutoEncodingVariationalBayes2013}.
But the variational bound also has an interesting interpretation in terms of compression / minimum description length \cite{chenVariationalLossyAutoencoder2016, gregorDeepAutoRegressiveNetworks2013, hintonKeepingNeuralNetworks, AutoencodersMinimumDescription, honkelaVariationalLearningBitsBack2004}.
Under this interpretation, the first term of the rhs of Eq. \ref{eq:bound} measures the expected number of bits required to encode $\mbf{x}$ given that we know a sample $\mbf{z} \sim q(\mbf{z} | \mbf{x})$.
More specifically, one can derive a code for $\mbf{x}$ from the decoder distribution $p(\mbf{x} | \mbf{z})$, which assigns roughly $-\log p(\mbf{x} | \mbf{z})$ bits to $\mbf{x}$ \cite{coverElementsInformationTheory2006}.
Averaged over $q$, one obtains the first term of the VAE loss (Eq. \ref{eq:bound}).

We note that in lossy compression, we do not actually encode $\x$ using $p(\x | \z)$, which would allow lossless reconstruction.
Instead, we only send $\z$ and hence refer to the first loss term as the distortion.


The second term of the bound (the KL) is related to the cost of coding the latents $\z$ coming from the encoder $q(\z | \x)$ using an optimal code derived from the prior $p(\z)$.
Such a code will use about $-\log p(\z)$ bits to encode $\z$.
Averaging over the encoder $q(\z | \x)$, we find that the average coding cost is equal to the cross-entropy between $q$ and $p$:
\begin{equation}
    E_{q(\mbf{z} | \mbf{x})}[-\log p(\mbf{z})]
    = \operatorname{CE}[q(\mbf{z}|\mbf{x}) | p(\mbf{z})].
\end{equation}


The cross-entropy is related to the KL via the relation $\operatorname{KL}[q|p] = \operatorname{CE}[q|p] - \operatorname{H}[q]$, where $\operatorname{H}[q]$ is the entropy of the encoder $q$.
So the $\operatorname{KL}$ measures the coding cost, except that there is a discount worth $\operatorname{H}[q]$ bits: randomness coming from the encoder is free.
It turns out that there is indeed a scheme, known as bits-back coding, that makes it possible to transmit $\z \sim q(\z|\x)$ and get $\operatorname{H}[q]$ bits back, but this scheme is difficult to implement in practice, and can only be used in lossless compression 
\cite{hintonKeepingNeuralNetworks}.

Since we cannot use bits-back coding for lossy compression, the cross-entropy provides a more suitable loss than the $\KL$.
Moreover, when using discrete latents, the entropy $\operatorname{H}[q]$ is always non-negative, so we can add it to the rhs of Eq. \ref{eq:bound} and obtain a valid bound.
We thus obtain the rate-distortion loss
\begin{equation}
    \label{eq:rd_loss}
    \mathcal{L}(\mbf{x}) = E_{q(\mbf{z} | \mbf{x})}[-\log p(\mbf{x} | \mbf{z}) - \beta \log p(\mbf{z})],
\end{equation}
where $\beta$ is a rate-distortion tradeoff parameter.

Since the cross-entropy loss does \emph{not} include a discount for the encoder entropy, there is a pressure to make the encoder more deterministic.
Indeed, for a fixed $p(\mbf{z})$ and $p(\mbf{x}|\mbf{z})$, the optimal solution for $q(\mbf{z}|\mbf{x})$ is a deterministic (``one hot'') distribution that puts all its mass on the state $\z$ that minimizes $-\log p(\x | \z) - \beta \log p(\z)$.

For this reason, we only consider deterministic encoders in this work.
When using deterministic encoders, the rate-distortion loss (Eq. \ref{eq:rd_loss}) is equivalent to the variational bound (Eq. \ref{eq:bound}), because (assuming discrete $\z$), we have $\operatorname{H}[q] = 0$ and hence $\operatorname{KL}[q|p] = \operatorname{CE}[q|p]$.



Finally, we note that limiting ourselves to deterministic encoders does not lower the best achievable likelihood, assuming a sufficiently flexible class of prior and likelihood.
Indeed, given \emph{any} fixed deterministic encoder $q$, we can still achieve the maximum likelihood by setting $p(\z) = \sum_{\x} p(\x) q(\z | \x)$ and $p(\x | \z) \propto p(\x) q(\z | \x)$, where $p(\x)$ is the true data distribution.

%% file: methods.tex
\begin{figure}[t!]
	\centering	
    \includegraphics[width=0.95\linewidth]{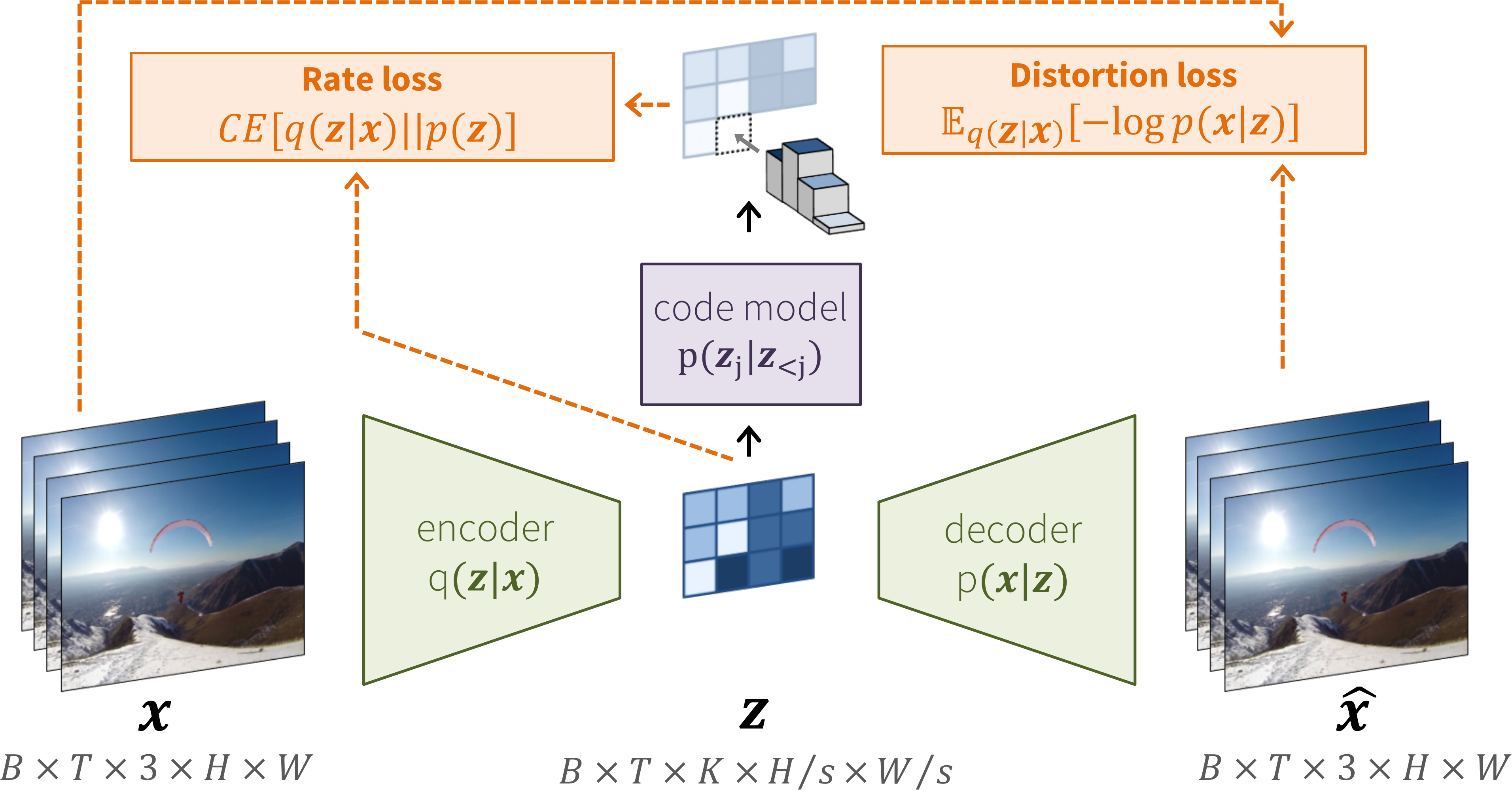}
    \caption{Training Rate-Distortion autoencoders. The rate loss is a measure for the expected coding cost, under the autoregressive code model, while the distortion loss expresses the reconstruction error.
    }
    \label{fig:model-training}
\end{figure}

In the previous section, we have outlined the general compression framework using rate-distortion autoencoders.
Here we will describe the specific models we use for encoder, code model, and decoder, as well as the data format, preprocessing, and loss functions. 

\subsection{Preprocessing}
Our model processes chunks of video $\x$ of shape $T \times C \times H \times W$, where $T=8$ denotes the number of frames, $C$ denotes the number of channels (typically $C=3$ for RGB), and $H, W$ are the height and width of a crop, which we fix to $160$ pixels in all of our experiments.
The RGB values are not scaled, \ie, they always lie in $\{0, 1, \ldots, 255\}$.

\subsection{Autoencoder}
\label{sec:autoencoder}
The encoder takes as input a chunk of video $\x$ and produces a discrete latent code $\mbf{z}$.
If the input has shape $T \times C \times H \times W$, the latent code will have shape $T \times K \times H/s \times W/s$, where $K=32$ is the number of channels in the latent space, and $s = 8$ is the total spatial stride of the encoder (so the latent space has spatial size $H/s=W/s = 160/8 = 20$).
We do not use stride in the time dimension.

The encoder and decoder are based on the architecture presented by \cite{mentzerConditionalProbabilityModels2018}, which in turn is based on the architecture presented in \cite{theisLossyImageCompression2017}.
The encoder and decoder are both fully convolutional models with residual connections \cite{heDeepResidualLearning2016}, batchnorm \cite{ioffeBatchNormalizationAccelerating2015}, and ReLU nonlinearities.
In the first two convolution layers of the encoder, this model uses filter size $5$ and stride $2$. 
The remaining layers are $5$ residual blocks with two convolution layers per block, filter size $3$, $128$ channels, batchnorm, and ReLU nonlinearities.
The final layer is a convolution with filter size $3$, stride $2$, and $32$ output channels.
The decoder is the reverse of this, and uses transposed convolutions instead of convolutions.
More details on the architecture can be found in the supplementary material.

We will evaluate two versions of this autoencoder: one with 2D convolutions applied to each frame separately, and one with 3D spatio-temporal convolutions.
To apply the 2D model to a video sequence, we simply fold the time axis into the batch axis before running the 2D AE.


The encoder network first outputs continuous latent variables $\tilde{\z}$, which are then quantized.
The quantizer discretizes the coordinates of $\tilde{\z}$ using a learned codebook consisting of $L$ centers, $\mc{C} = \{c_{1}, \ldots, c_{L}\}$, where $c_{l} \in \mb{R}$.
In the forward pass, we compute $z_{j} = \arg\min_i |\tilde{z}_{j} - c_i|$ (where $j=(t,c,h,w)$ is a four dimensional multi-index).
As a probability distribution, this corresponds to a one-hot $q(z_{j} | \x)$ that puts all mass on the computed value $z_{j}$.
Because the argmin is not differentiable, we use the gradient of a softmax in the backward pass, as in \cite{bengioEstimatingPropagatingGradients2013,mentzerConditionalProbabilityModels2018}.
We found this approach to be stable and effective during training.


On the decoder side, we replace $z_{j} \in \{1, \ldots, L\}$ by the corresponding codebook value $c_{z_{j}}$, to obtain an approximation of the original continuous representation $\tilde{\z}$.
The resulting vector is then processed by the decoder to produce a reconstruction $\hat{\x}$.
In a standard VAE, one might use $\hat{\x}$ as the mean of a Gaussian likelihood $p(\x|\z)$, which corresponds to an L2 loss: $-\log p(\x|\z) \propto \|\x - \hat{\x}\|^2 + const$.
Instead, we use the MS-SSIM loss (discussed in Sec. \ref{sec:loss}), which corresponds to the unnormalized likelihood of the Boltzmann distribution, $\ln p(\mathbf{x}|\mathbf{z}) = \textrm{ms-ssim}(\mathbf{x},\hat{\x}) - \ln C$, where $\ln C$ is the log-partition function treated as a constant, because it better reflects human subjective judgments of similarity.

\begin{figure}[t!]

  \begin{center}
  \resizebox{.45\textwidth}{!}{%
     \subfloat[Unconditional]{
	   \includegraphics[height=3.1cm]{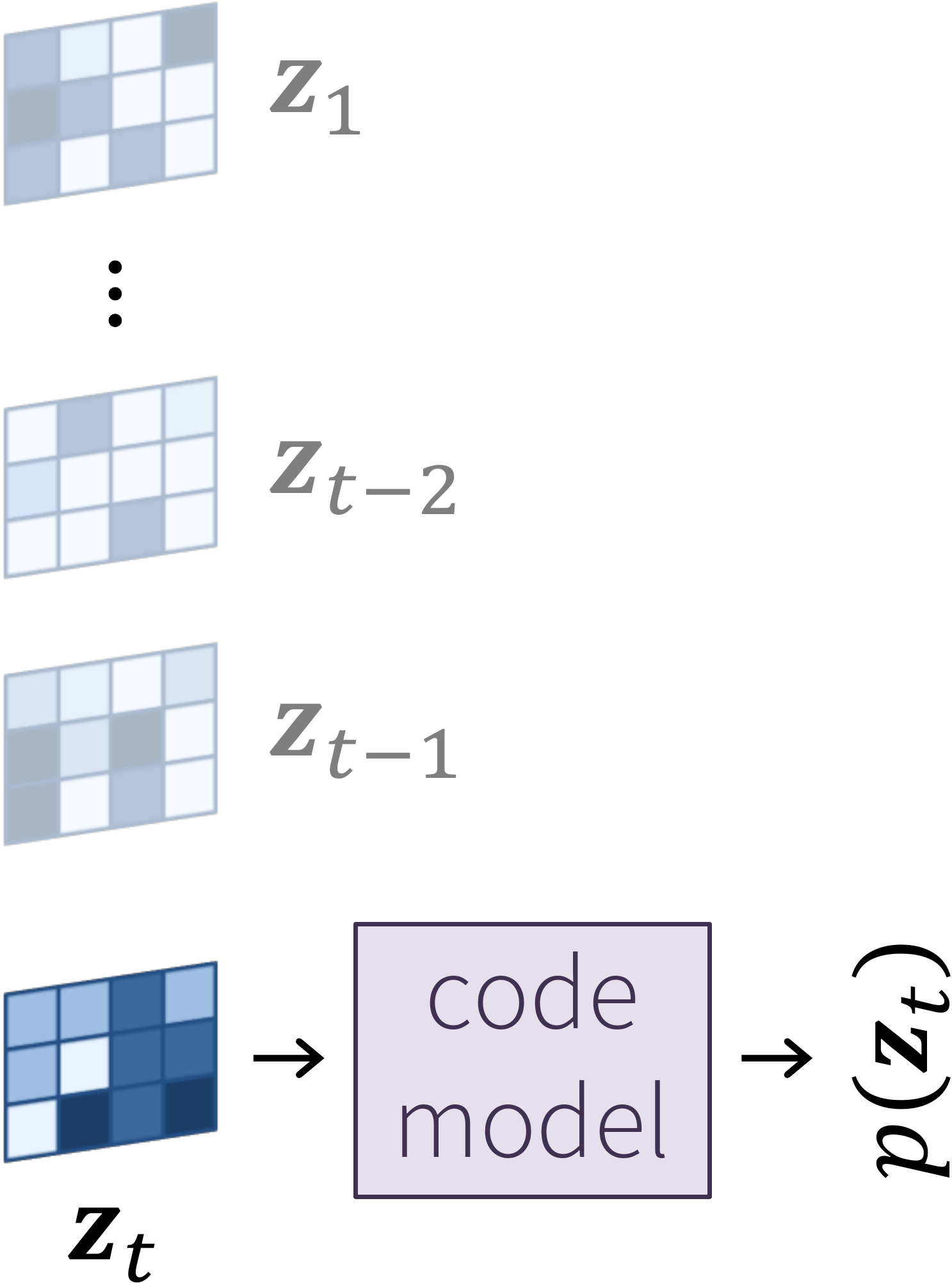}
     \;} 
     \subfloat[Frame-conditioned]{
	   \includegraphics[height=3.1cm]{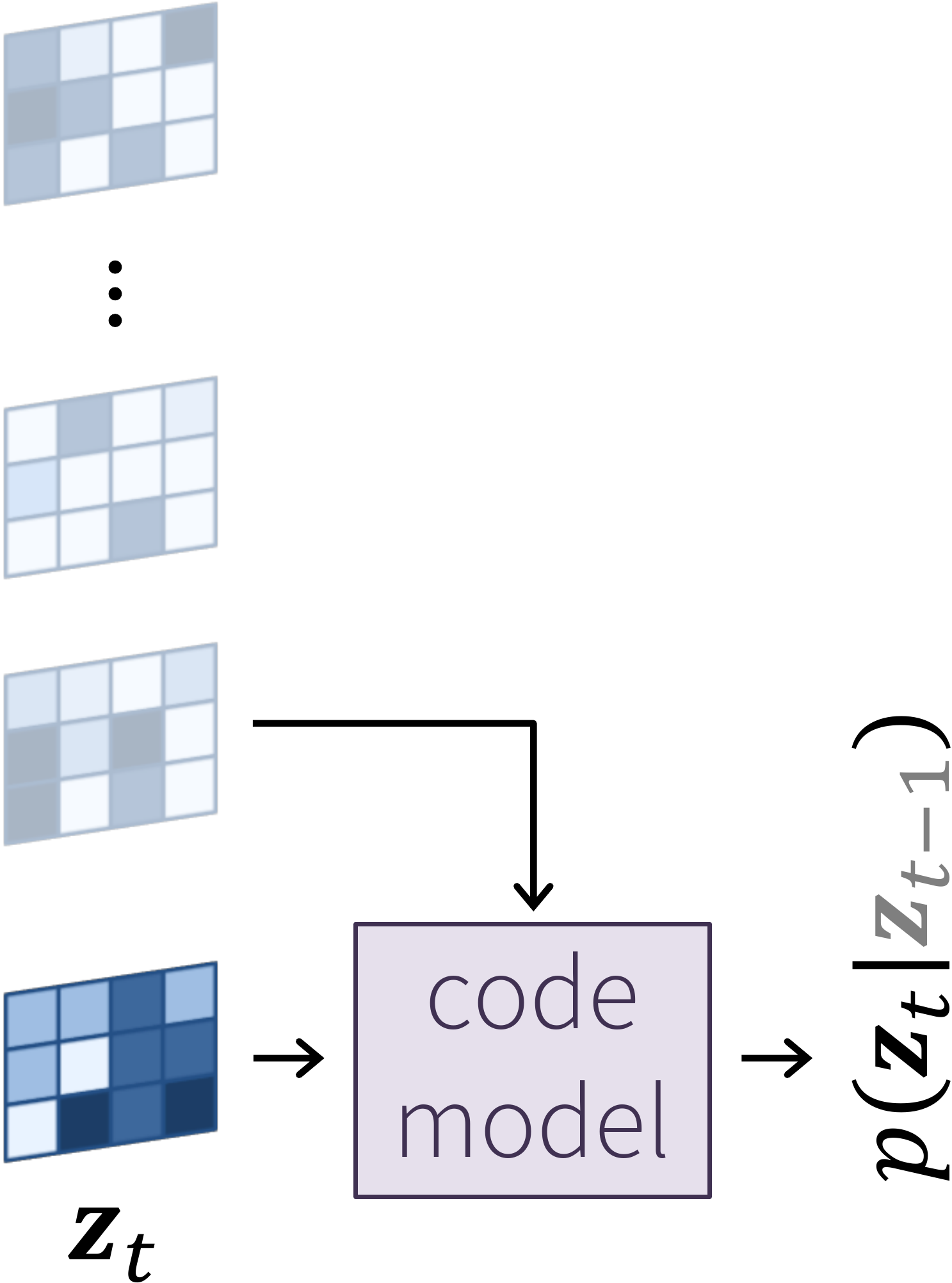}
    	\; } 
    	\subfloat[GRU-conditioned]{
	   \includegraphics[height=3.1cm]{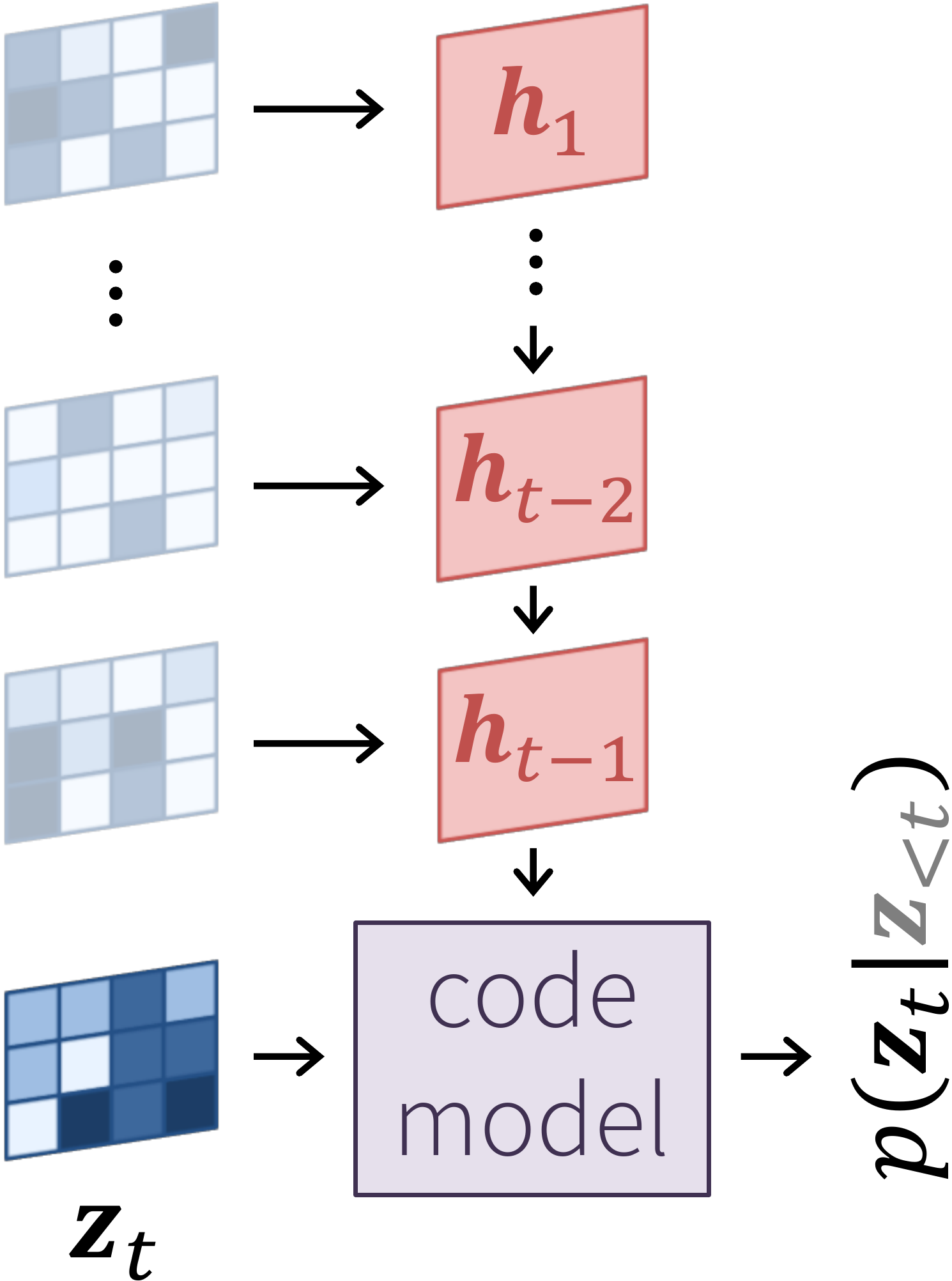}
	}
  }
  \end{center}
   \caption{Proposals for temporal conditioning of prior.}
    \label{fig:priors}
\end{figure}

\begin{figure*}[t!]
    \centering
    \subfloat[AVC/H.264  (0.037 BPP)] {
    \includegraphics[scale=0.31]{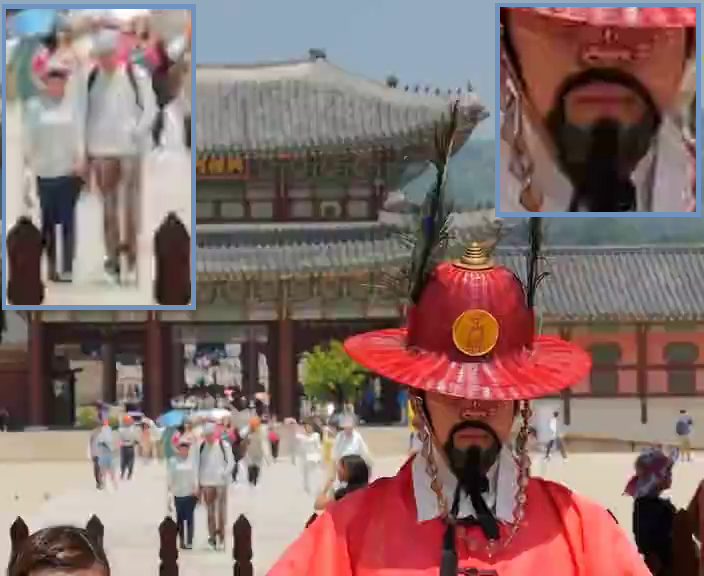}}
    \subfloat[HEVC/H.265 (0.036 BPP)] {
    \includegraphics[scale=0.31]{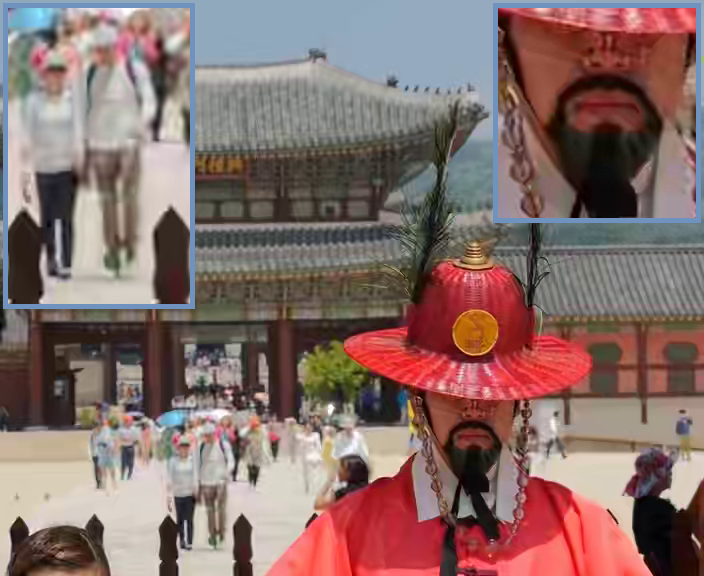}}
    \subfloat[Our model (0.037 BPP)] {
    \includegraphics[scale=0.31]{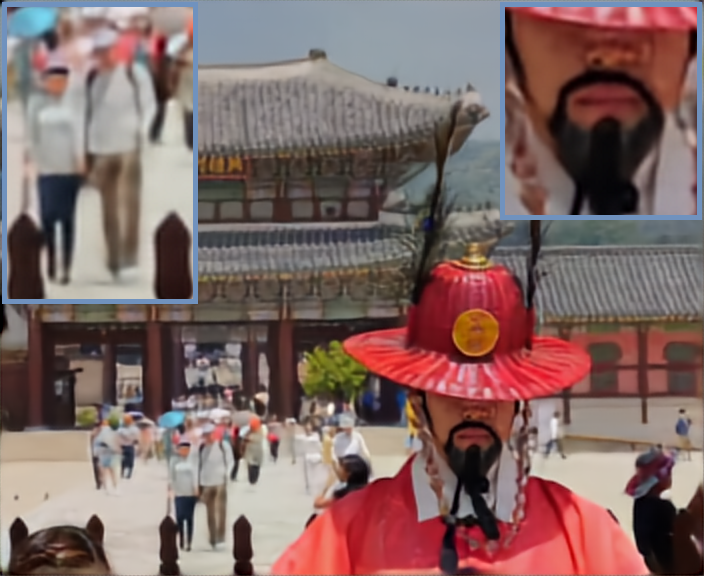}}
    \caption{Compression results for the state-of-the-art traditional codecs, AVC/H.264 and HEVC/H.265, and our proposed model. On a similar bitrate, our model approachs these codecs while generatinng less artifacts.
    }
    \label{fig:qualitative}
\end{figure*}

\subsection{Autoregressive Prior}

Instead of naively storing / transmitting the latent variables $\z$ using $D \log_2 L$ bits (for a $D$-dimensional latent space with $L$ states per variable), we encode the latents using the prior $p(\z)$ in combination with adaptive arithmetic coding.
For $p(\z)$, we use a gated PixelCNN \cite{vandenoordConditionalImageGeneration2016a} over individual latent frames, optionally conditioned on past latent frames as in video pixel networks \cite{kalchbrennerVideoPixelNetworks2016}.
In Figure \ref{fig:priors}, we illustrate the three priors considered in this paper.

In the simplest case, we model each frame independently, i.e. $p(\z) = \prod_t p(z_t)$, where a latent frame $z_t$ is modelled autoregressively as $p(z_t) = \prod_{i} p(z_{t, i} | z_{t,<i})$ by the PixelCNN.
Here $i = (c, h, w)$ denotes a 3D multi-index over channels and spatial axes, and $z_{t,{<i}}$ denotes the elements that come before $i$ in the autoregressive ordering. 

A better prior is obtained by including temporal dependencies (Figure \ref{fig:priors}b).
In this model, the prior is factorized as $p(\z) = \prod_t p(z_t|z_{t-1})$, where $p(z_t|z_{t-1}) = \prod_{i} p(z_{t,i} | z_{t,<i}, z_{t-1})$.
Thus, the prediction for pixel $i=(c,h,w)$ in latent frame $t$ is based on previous pixels in the same frame, as well as the whole previous frame $z_{t-1}$.
The dependence on $z_{t,<i}$ is mediated by the masked convolutions of the PixelCNN architecture, whereas the dependence on the previous frame $z_{t-1}$ is mediated by additional conditioning connections added to each layer, as in the original Conditional PixelCNN \cite{vandenoordConditionalImageGeneration2016a}.

Conditioning on the previous frame may be limiting if long-range temporal dependencies are necessary.
Hence, we also consider a model where a recurrent neural network (Gated Recurrent Units, GRU) summarizes all relevant information from past frames.
The prior factorizes as $p(\z) = \prod_t p(z_t | z_{<t})$ with $p(z_t | z_{<t}) = \prod_i p(z_{t,i} | h_{t-1}, z_{t,<i})$, where $h_{t-1}$ is the hidden state of a GRU that has processed latent frames $z_{<t}$. 
As in the frame-conditional prior, in the GRU-conditional prior, the dependency on $z_{t,<i}$ is mediated by the causal convolutions of the PixelCNN, and the dependency on $h_t$ is mediated by conditioning connections in each layer of the PixelCNN.


\subsection{Loss functions, encoding, and decoding}
\label{sec:loss}

To measure distortion, we use the Multi-Scale Structural Similarity (MS-SSIM) loss \cite{wangMultiScaleStructuralSimilarity2003}.
This loss gives a better indication of the subjective similarity of $\hat{\x}$ and $\x$ than a simple L2 loss, and has been popular in (learned) image compression.
To measure rate, we simply use the log-likelihood $-\log p(\z)$ where $\z$ is produced by the encoder deterministically.
The losses are visualized in Figure \ref{fig:model-training}.




To encode a chunk of video $\x$, we map it through the enncoder to obtain latents $\z$.
Then, we go through the latent variables one by one, and make a prediction for the next latent variable using the autoregressive prior $p(z_j | z_{<j})$.
We then use an arithmetic coding algorithm to obtain a bitstream $b_j = \operatorname{ENC}(z_j, p(z_j | z_{<j}))$ for the $j$-th variable.
The expected length of $b_j$ is $-\log p(z_j | z_{<j}))$.

To decode, we take the bitstream $b_j$ and combine it with the prediction $p(z_j | z_{<j})$ to obtain $z_j = \operatorname{DEC}(b_j, p(z_j | z_{<j})$.
Once we have decoded all latents, we pass them through the decoder of the AE to obtain $\hat{\x}$.


%% file: experiments.tex
\subsection{Dataset}
\label{sec:dataset}

\textbf{Kinetics}~\cite{carreira2017quo} We use videos with a width and height greater than $720px$, which results in $98,944$ videos as our training set. We only use the first $16$ frames for training. The resulting dataset has about $1.6m$ frames, which is sufficient for training our model, though larger models and datasets will likely result in better rate/distortion (at the cost of increased computational cost during training and testing).

\textbf{Ultra Video Group}~\cite{UVG} UVG contains $7$ videos with $3,900$ frames in full HD resolution ($1920 \times 1080$). We use this dataset to compare with state-of-the-art.

\textbf{Standard Definition Videos} SDV contains $20$ videos with $\sim 40K$ frames of resolution $352 \times 288$. We use this dataset for ablation studies.

\textbf{Human Activity} contains $1257$ real-world videos of people in various everyday scenes, and is mostly used for human pose estimation and tracking in video. Following the standard partitions of the data, we use $1087$ and $170$ videos as train and test set for semantic compression experiments.

\textbf{Dynamics} is an internal dataset containing ego-view video from a car driving on different highways at different times of day. The full dataset consists of $5$ clips taken at different dates, times, and locations. We use $4$ clips of $20$ minutes each (120k frames) as train set, and use the fifth clip of $14$ minutes (25k frames) as test sequence.

\textbf{Berkeley MHAD}~\cite{ofliBerkeleyMHADComprehensive2013} contains videos of human actions, recorded by four multi-view cameras. We use this dataset for multi-modal compression experiments. We use all four video streams from the first quad-camera, each of which records the same scene from a slightly shifted vantage point. The MHAD dataset contains $11$ actions each performed by $12$ participants, with $5$ repetitions per participant.
We use the first $4$ repetitions for training, and the last one for testing.

Kinetics, Dynamics and Human Activity are only available in compressed form, and hence contain compression artifacts.
In order to remove these artifacts, we downscale videos from these datasets so that the smallest side has length $256$, before taking crops.
For uncompressed datasets (UVG, SDV, and MHAD), we do not perform downscaling.

\subsection{Training}
\label{sec:setup}

We train all of our models with batchsize $32$, using the Adam optimizer \cite{kingmaAdamMethodStochastic2015} with learning rate $10^{-4}$ (decaying with $\gamma=0.1$ every $40$ epochs) for a total of $N=100$ epochs.
Only for the Kinetics dataset, which is much larger, we use $10$ epochs and learning rate decay every $4$ epochs.

We use MS-SSIM (multi-scale structural similarity) as a distortion loss, and the cross-entropy as a rate loss.
In order to obtain rate-distortion curves, we train separate models for beta values $\beta \in \{0.1, 0.3, 0.5, 0.7\}$ (unless stated otherwise), and report their rate/distortion score.

%

%
\begin{figure}[t!]
    \centering
    \includegraphics[scale=0.42]{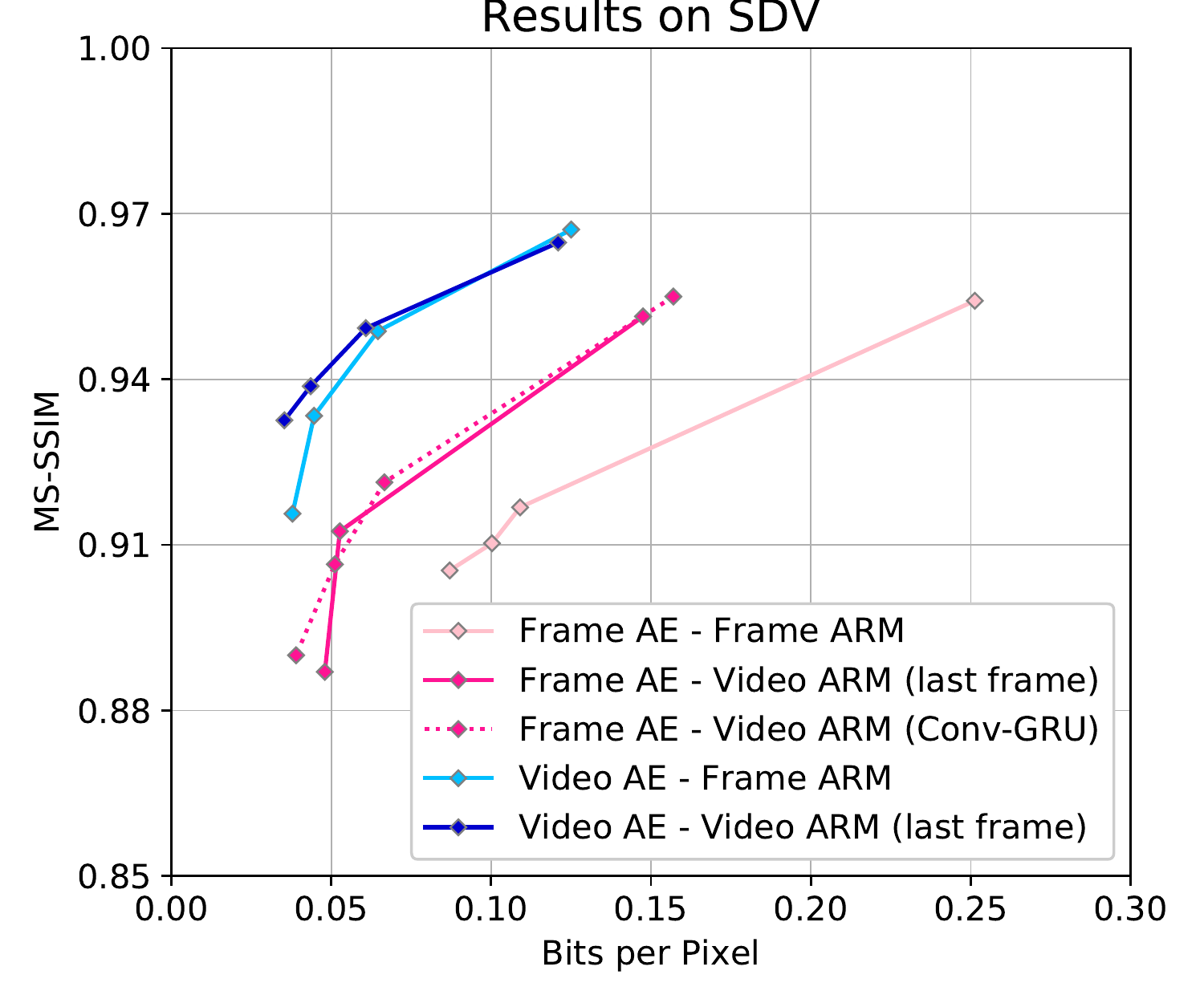}
    \caption{Ablation experiments. The both autoencoder and prior exploit temporal dependencies, in pixel and latent space respectively, to improve video compression.}
    \label{fig:ablation_experiments}
\end{figure}

\subsection{Ablation studies}
\label{secab:ablation}

We evaluate several AE and prior design choices as discussed in Section~\ref{sec:methods}. 
Specifically, we compare the use of 2D and 3D convolutions in the autoencoder, Frame AE and Video AE respectively, as well as three kinds of priors: a 2D frame-based ARM that does not exploit temporal dependencies (Frame ARM), an ARM conditioned on the previous frame (Video ARM-last frame), and one conditioned on the output of a Conv-GRU (Video ARM-Conv-GRU). We train each model on Kinetics and evaluate on SDV.

The results are presented in Figure~\ref{fig:ablation_experiments}.
The results show that conditioning the ARM on the previous frame yields a substantial boost over frame-based encoding, particularly when using a frame AE.
Introducing a Conv-GRU only marginally improves results compared to conditioning on the last frame only.

We also note that using the 3D autoencoder is substantially better than using a 2D autoencoder, even when a video prior is not being used.
This suggests that the 3D AE is able to produce latents that are temporally decorrelated to a large extent, so that they can be modelled fairly effectively by a frame AE.
The difference between 2D and 3D AEs is substantially bigger than the difference between 2D and 3D priors, so in applications where a few frames of latency is not an issue, the 3D AE is to be preferred, and can reduce the burden on the prior.

For the rest of the experiments, we will use the best performing model: the Video AE + Video ARM (last frame).

\subsection{Comparison to state of the art}
\label{sec:compare_sota}
\begin{figure}[t!]
    \centering
    \includegraphics[scale=0.42]{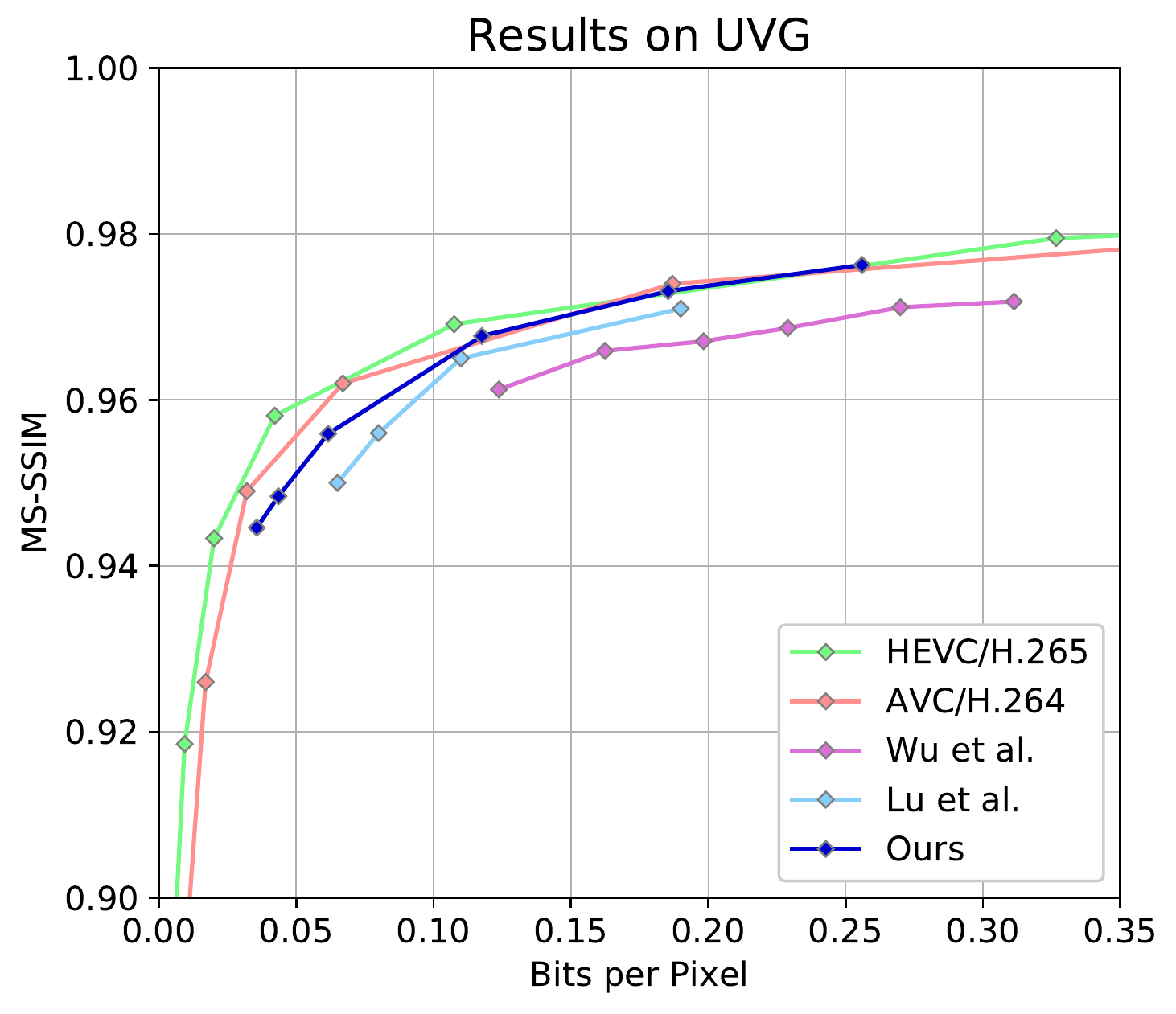}
	\caption{Comparison to the state-of-the-art traditional and learned codecs. Our proposal outperforms the learned counterparts and approaches AVC/H.264 and HEVC/H.265 evaluated in their default setting.}    
    \label{fig:standard_compression_uvg}
\end{figure}
We benchmark our method against the state-of-the-art traditional and learned compression methods on UVG standard test sequences.
We compare against classical codecs AVC/H.264 and HEVC/H.265, as well as the recent learned compression methods presented by \cite{luDVCEndtoendDeep2018} and \cite{wuVideoCompressionImage2018}.
For the classical codecs, we use the default FFmpeg settings, without imposing any restriction, and only vary the CRF setting to obtain rate/distortion curves.
For the other learned compression methods, we use the results as reported in the respective papers.
For our method, we use 6 different $\beta$ values, namely, $0.03, 0.05, 0.1, 0.3, 0.5, 0.7$.

Figure \ref{fig:standard_compression_uvg} shows that our method consistently outperforms other learned compression methods, and is approaching the performance of classical codecs, particularly in the $0.10 - 0.25$ bpp range.
We note that in some previous works, learned compression was shown to outperform classical codecs, when the latter are evaluated under restricted settings by limiting the inter-frame compression to only few frames, \ie by setting \emph{GOP} flag to $12$. The results under restricted setting are reported in supplementary materials.

\begin{figure*}[t!]
    \centering
    \subfloat[Semantic Compression] {
        \includegraphics[scale=0.38]{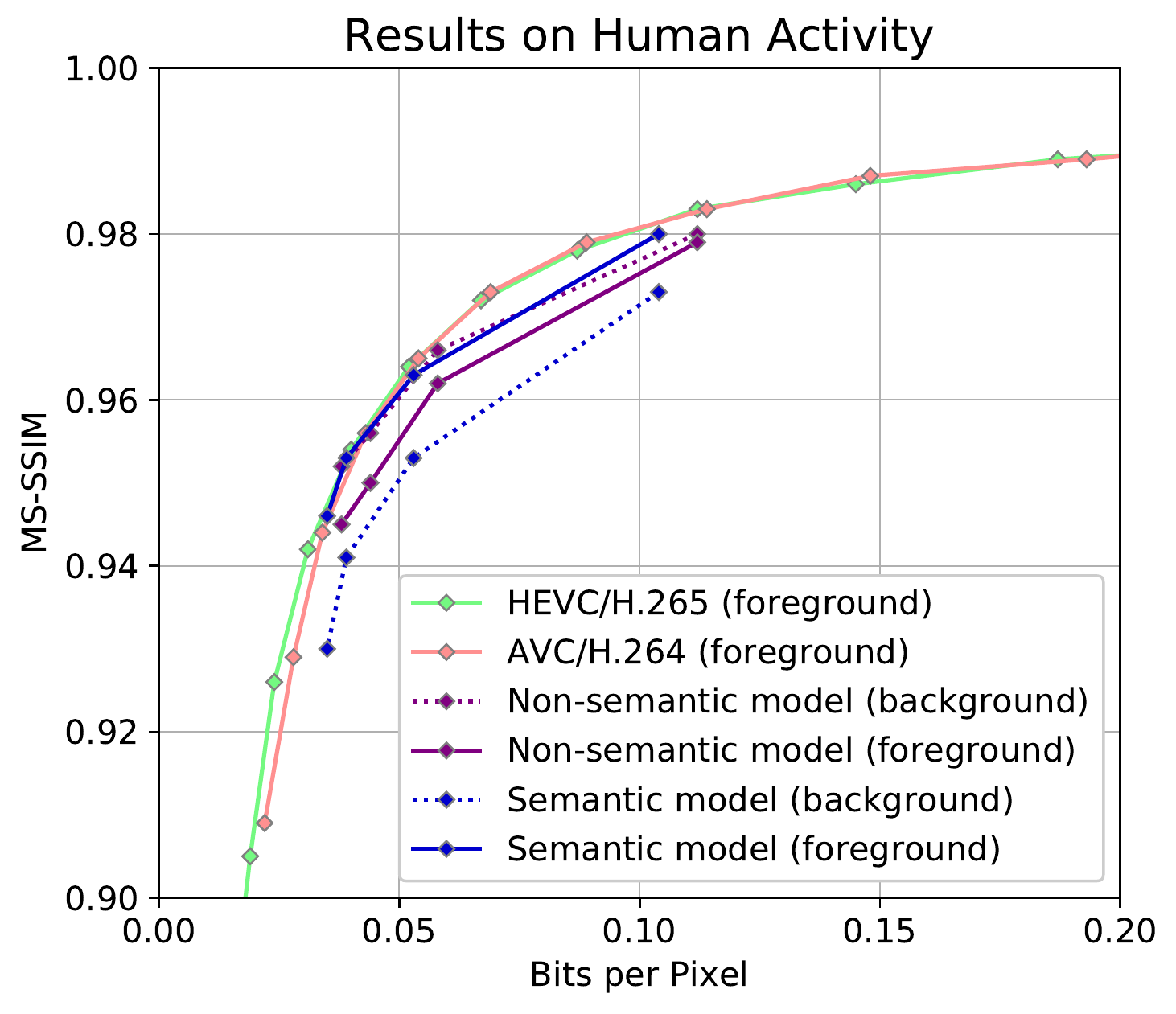}
    		\label{fig:semantic_compression}
    	}    
    \subfloat[Adaptive Compression] {
    		\includegraphics[scale=0.38]{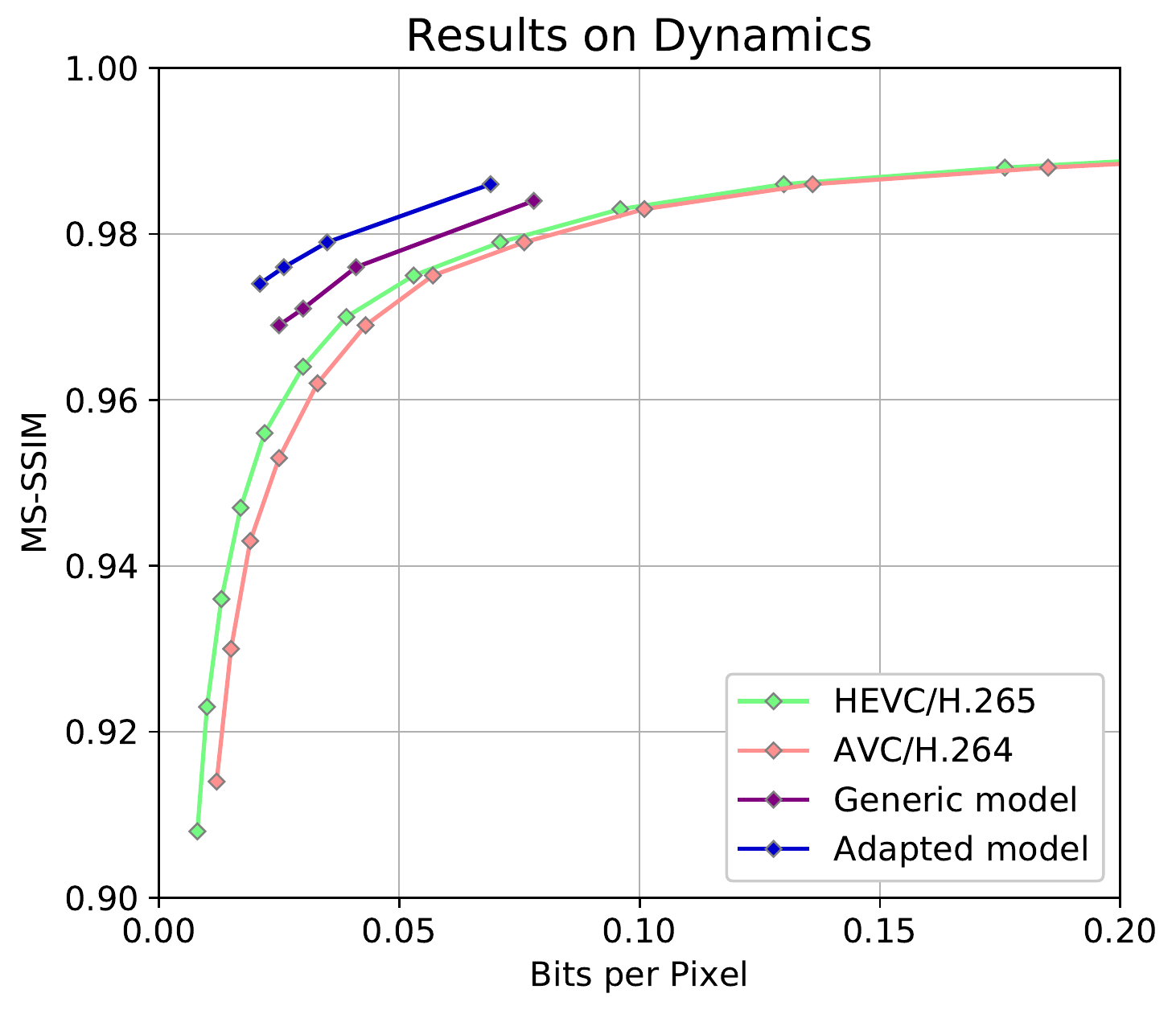}
		\label{fig:adaptive_compression}    
    }
    \subfloat[Multimodal Compression] {
    		\includegraphics[scale=0.38]{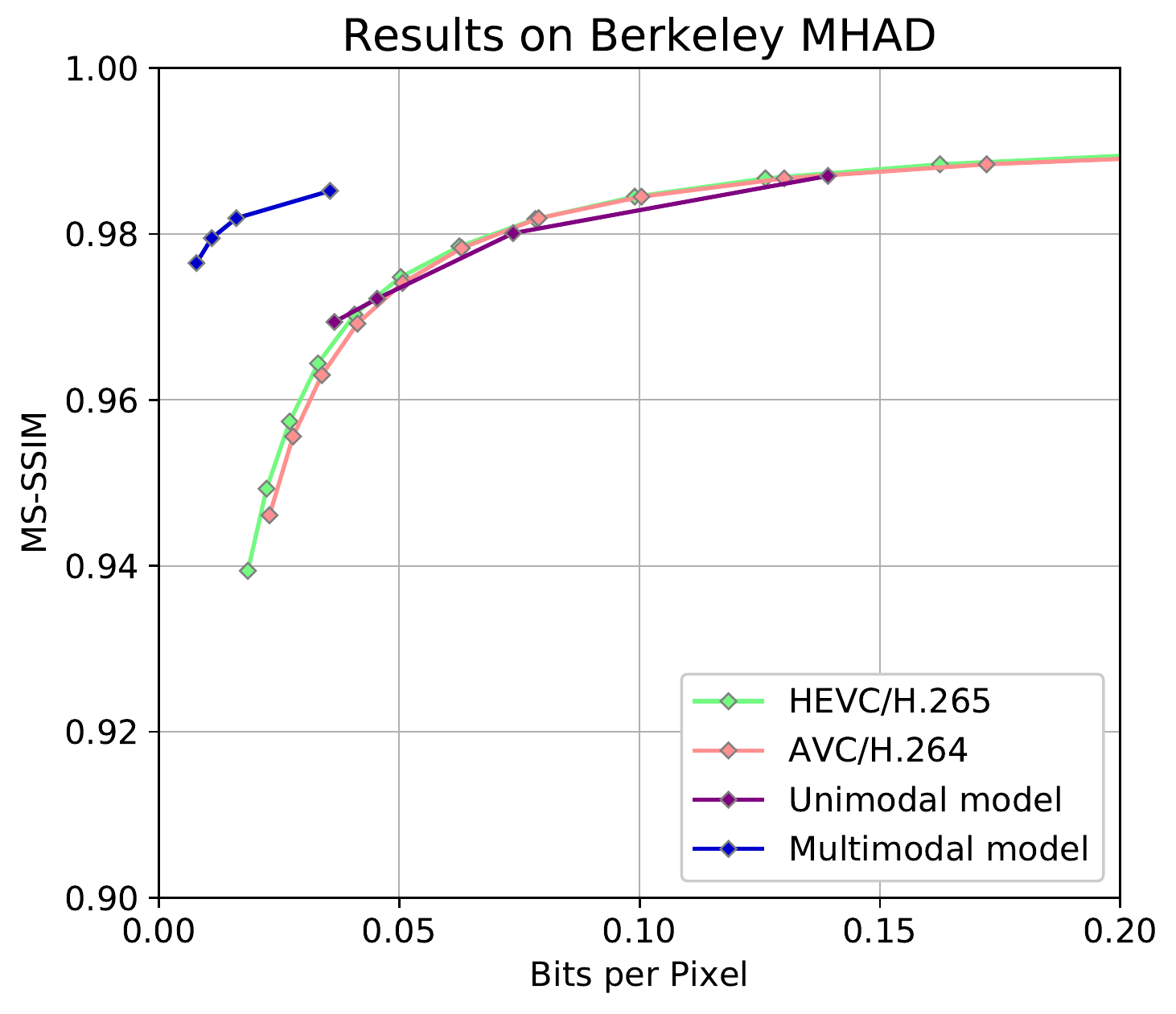}
    		\label{fig:multimodal_compression}
    	}    	
    \label{fig:triple_exp}
   	\caption{Three extensions of our model that demonstrate the benefits of learned over classical approaches to compression.}
\end{figure*}

\subsection{Semantic Compression}
The perceived quality of a compressed video depends more on how well salient objects are reconstructed, and less on how well non-salient objects are reconstructed.
For instance, in video conferencing, it is more important to preserve details of faces than background regions.
It follows that better subjective quality can be achieved by allocating more bits to salient / foreground objects than to non-salient / background objects.

Developing such a task-tuned video codec requires a semantic understanding of videos.
This is difficult to do with classical codecs as it would require distinguishing foreground and background objects. 
For learned compression methods, the asymmetry is easily incorporated by using different weights for the rate/distortion losses for foreground (FG) and background (BG) objects, assuming that ground-truth FG/BG annotations are available during training.

In this experiment, we study the semantic compression of the \emph{person} category. The groundtruth person regions are extracted using a Mask R-CNN \cite{heMaskRCNN2017} trained on COCO images. We use bounding boxes around the objects, but the approach is applicable to segmentation masks without any modification required.
The detected person regions are converted to a binary mask and used for training.

The MS-SSIM loss is a sum over scales of the SSIM loss.
The SSIM loss computes an intermediate quantity called the similarity map, which is usually aggregated over the whole image.
Instead, we aggregate these maps separately for foreground and background, where the FG and BG mask at a given scale is obtained from the high-resolution mask by average pooling.
We then sum the FG and BG components over each scale, and multiply the resulting FG and BG losses by separate weights $\alpha$ and $1 - \alpha$, respectively. We set the $\alpha$ to $0.95$ in our experiments.

The rate loss is a sum of $-\log p(z_i | z_{<i})$, so we can multiply each term with a foreground/background weight.
Each latent covers an $8 \times 8$ region of pixels, thus, we need to aggregate the pixel-wise labels to obtain a label for each latent.
We do this by average pooling the FG/BG mask over $8 \times 8$ regions to obtain a weight per latent position which we multiply with the rate loss at that position.

The results are shown in Figure~\ref{fig:semantic_compression}. 
We observe that in the non-semantic model, BG is reconstructed more accurately than FG at a fixed average bitrate. The same behavior is observed for classical codecs as reported in supplementary materials. The worse reconstruction of FG is not surprising because person regions usually contain more details compared to the more homogeneous background regions. However, when using semantic loss weighting, the relation is reversed. Semantic loss weighting leads to an improvement in MS-SSIM score for FG at the expense of MS-SSIM score for BG. It demonstrates the effectiveness of learned video compression in incorporating semantic understanding of video content into compression. We believe that it opens up novel video compression applications which have not been feasible with classical codecs.

%

\begin{figure*}[t!]
    \centering
    \includegraphics[scale=0.40]{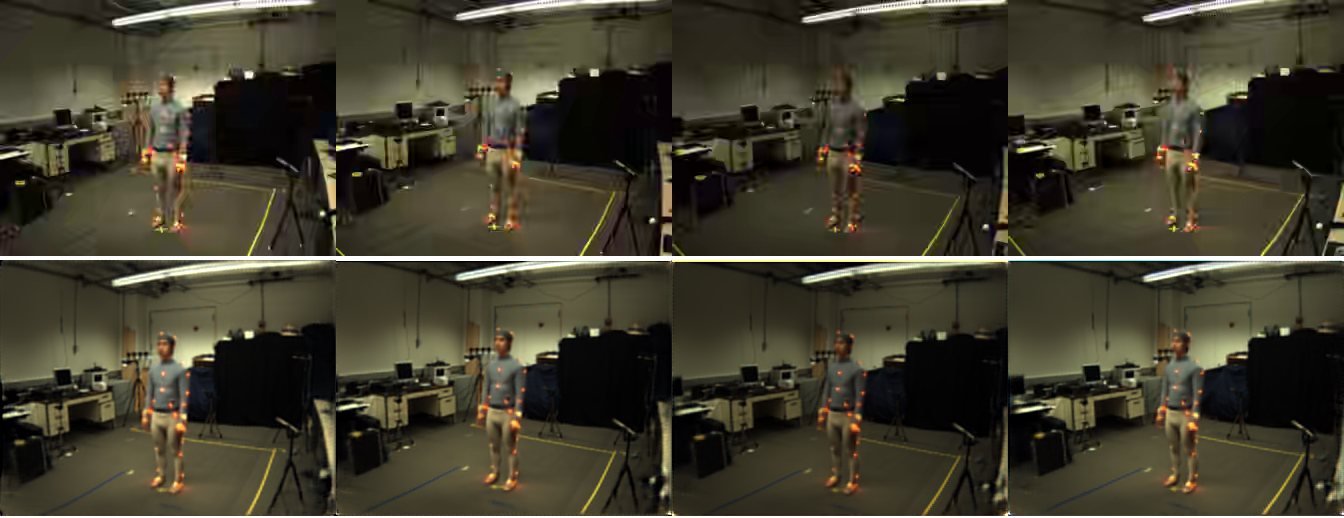}
    \caption{Multimodal compression results for HEVC/H.265 (top) and our proposal (bottom). By utilizing the redundancies between different views of a quad camera (columns), our model achieves a significantly better reconstruction while using $5 \times$ less bits ($0.007$ vs $0.035$ BPP).}
    \label{fig:qualitative_multimodal}
\end{figure*}

\subsection{Adaptive Compression}

Classical codecs are optimized for good performance across a wide range of videos.
However, in some applications, the codec is used on a distribution of lower entropy videos, \emph{i.e.} scenes with predictable types of activities. For example, a security camera placed at a fixed location and viewpoint will produce a very predictable video. In this experiment we show that learned compression models can utilize the lower entropy videos by simply being finetuned on them, which is difficult to do with classical codecs. 

In this experiment, we show that by finetuning a learned compression model on the Dynamics dataset, substantial improvements in compression can be achieved. Figure~\ref{fig:adaptive_compression} compares the classical codecs with our \emph{generic model} as well as the \emph{adapted model}. The generic model is trained on a generic training set from Kinetics. The adapted model takes a pretrained generic model and finetunes it on videos of a similar domain. The results show that our generic method outperforms the classical codecs on this dataset, and the adapted method shows even better performance.

This experiment indicates a great practical potential of learned compression models. Finetuning a compression model allows to maintain high reconstruction quality with substantially lower compression rate, while the model could be transferred from a generic compression model.

\subsection{Multimodal Compression}

Classical codecs are designed for typical videos captured by monocluar color cameras.
When other modalities are included, such as depth, stereo, audio, or spectral imaging sensors, classical codecs are often not applicable or not able to exploit dependencies which exist between various modalities.
Developing a codec for every new modality is possible, but very expensive considering the amount of engineering work involved in designing classical codecs.
Using our learned compression method, however, adding new modalities is as easy as retraining the model on a new dataset with minimal modifications required.

In this experiment, we adapt our learned compression method to compress videos of human actions recorded by quad (four view) cameras from MHAD dataset. We compare four methods: \emph{AVC/H.264} and \emph{HEVC/H.265}, as well as a learned \emph{unimodal model} and a learned \emph{multimodal model}.
The unimodal model is trained on the individual video streams, and the multimodal model is trained on the channel-wise concatenation of the four streams.
The network architecture for the unimodal model and the multimodal model is the same as the one described in Section~\ref{sec:methods}, the only difference being that the multimodal model has more input channels ($4 \times 3$ vs $3$).

Interestingly, our approach retains more details than the classical codec (e.g., see the face of a person in Figure \ref{fig:qualitative_multimodal}) while obtaining $5$ times smaller BPP. The quantitative results, shown in Figure \ref{fig:multimodal_compression}, show that the multimodal compression model substantially outperforms all three baselines by utilizing the great amount of redundancy which exist between multiple data modalities.
This shows that without further tuning of the architecture or training procedure, our method can be applied to compress spatio-temporal signals from non-standard imaging sensors.

%% file: conclusion.tex

We have presented a video compression method based on variational autoencoders with a deterministic encoder.
Our theoretical analysis shows that in lossy compression, where bits-back coding cannot be used, deterministic encoders are preferred. 
Concretely, our model consists of an autoencoder and an autoregressive prior.
We found that 3D spatio-temporal autoencoders are very effective, and greatly reduce the need for temporal conditioning in the prior.
Our best model outperforms recent learned video compression methods without incorporating video-specific techniques like flow estimation or interpolation, and performs on par with the latest non-learned codec H.265 / HEVC. 

In addition, we have explicitly demonstrated the potential advantages of learned over non-learned compression, beyond mere compression performance.
In semantic compression, the rate and distortion losses are weighted by the semantics of the video content, giving priority to important regions, 
 resulting in better visual quality at lower bitrates in those regions.
In adaptive compression, a pretrained video compressor is finetuned on a specific dataset. 
With minimal engineering effort, this yields a highly effective method for compressing domain specific videos.
Finally, in our multi-modal compression experiments, we have demonstrated a dramatic improvement in compression performance, obtained simply by training the same model on a multi-modal dataset consisting of quad-cam footage.

%% file: supplementary.tex
\appendix
\FloatBarrier
\section{Images used in figures}
Video used in Figures \ref{fig:model-inference} and \ref{fig:model-training} by Ambrose Productions, and Figure \ref{fig:qualitative} by TravelTip. Both [CC BY-SA 3.0 https://creativecommons.org/licenses/by/3.0/legalcode], via YouTube.

\section{Architectural details}
In this section we detail the architecture and hyperparameters of the autoencoder and code-model that we use in our experiments. 

\subsection{Autoencoder}
\begin{figure}
    \centering
    \includegraphics[scale=0.35]{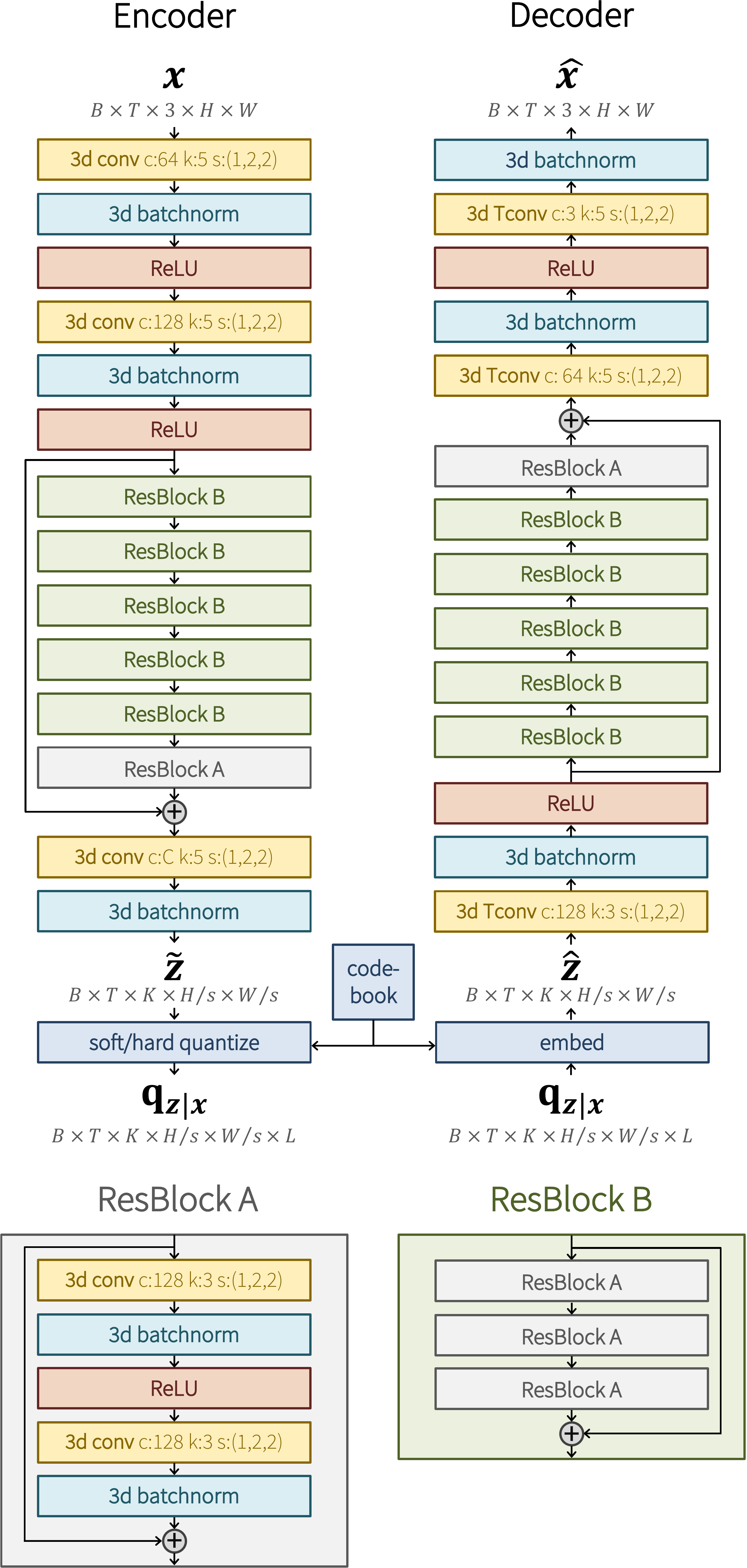}
    \vspace{5mm}
    \caption{Architecture of our autoencoder. \emph{Tconv} denotes transposed convolution. For (transposed) convolutional layers $c$ denotes the number of output channels, $k$ denotes the kernel size and $s$ denotes the stride. These are either expressed as $(x,y,z)$ triplets or as a single number that is used for each dimension. ``Same''-padding is used for all (transposed) convolution layers. $\q_{\z|\x}$ refers to the tensor of one-hot probabilities $ q(z_j = i | \x)$.
}
    \label{fig:model_architecture_autoencoder}
\end{figure}

As decribed in section \ref{sec:autoencoder}, we design our autoencoder by extending the (2D) model of \cite{mentzerConditionalProbabilityModels2018} to use 3D convolutions. The exact model is depicted in Figure \ref{fig:model_architecture_autoencoder}.

\subsubsection{Quantization}
As explained in Section \ref{sec:autoencoder}, the encoder network outputs continuous latent variables $\tilde{\z} \in \mb{R}^{B \times T \times K \times  H/s \times W /s}$, which are then quantized using a learned codebook $\mc{C} = \{c_{1}, \ldots, c_{L}\}$. 

Quantization involves computing $\q_{\z|\x} \in \mb{R}^{B \times T \times K \times  H/s \times W /s \times L}$ that  defines the probability for each codebook center $i$ at each position $j$ in $\z$ (note that this is one-hot over the $L$ axis).
We assume independence between all elements of $\z$ given $\x$, and we use the codebook distance to compute $\q_{\z|\x}$:

\begin{align}
q_{\z|\x}^{ij} &= q(z_j = i | \x) = \dfrac{ e^{-\tau |\tilde{z}_{j} - c_i|}  }{\sum_{k=1}^L e^{-\tau |\tilde{z}_{j} - c_{k}|} }
\end{align}

Where for notational simplicity, we are using a single index $j$ to index over all $(B,T,K,H,W)$ dimensions of $\tilde{\z}$.

Note that as $\tau \to \infty$,  $q(z_j=i| \x)$ will put more and more probability mass on  a single (the closest) center and will eventually be deterministic. This is desirable, as we want a deterministic encoder. In practice, we use a $\tau=10^{7}$ which we observe to always give us one-hot vectors for 32-bit precision floats. In the backward pass, we use the gradient of a softmax with a $\tau=1$ for numerical stability. 

On the decoder side, the one-hot probabilities $\q_{\z | \x}$ are embedded using the same codebook $\mc{C}$ to obtain the scalar tensor $\hat{\z}$ approximating $\tilde{\z}$ that is then decoded to predict $\x$.

\subsection{Autoregressive Code-Model}

The code model takes as input the one-hot probability tensor $\q_{\z|\x}$ output by the encoder, and predicts the probability for each entry $j$ in $\z$ in an autoregressive manner:

\begin{align}
p_{\z}^{ij} (\q_{\z | \x}) &= p(z_j = i | \z_{<j}) 
\end{align}

We use a 4 layer PixelCNN \cite{vandenoordConditionalImageGeneration2016a}  architecture with a kernel size of 5x5 and 8 hidden channels ($h=8$). We embed the one-hot probabilities of $\q_{\z|\x}$ using a learnable scalar embedding. We experimented with using the encoder codebook as the prior embedding, but we found that it did not make any difference in performance in practice.

\subsubsection{Conditioning}
For the frame-conditioned and GRU-conditioned model (see Figure \ref{fig:priors}), we inject the conditioning variable into each of the autoregressive blocks, right before applying the gated nonlinearity.

This conditioning input is a featuremap, and its number of channels should match the number of channels in the ARMBlock. The gated nonlinearity requires two times the number of hidden channels $h$. As there is a nonlinearity for both the horizontal and vertical stack, this would require $4h$ channels. Since the filters in PixelCNN are fully connected along the channel dimension, the required number of output channels for the conditioning featuremap is $(4h)K$. 

\begin{figure}
    \centering
    \includegraphics[scale=0.35]{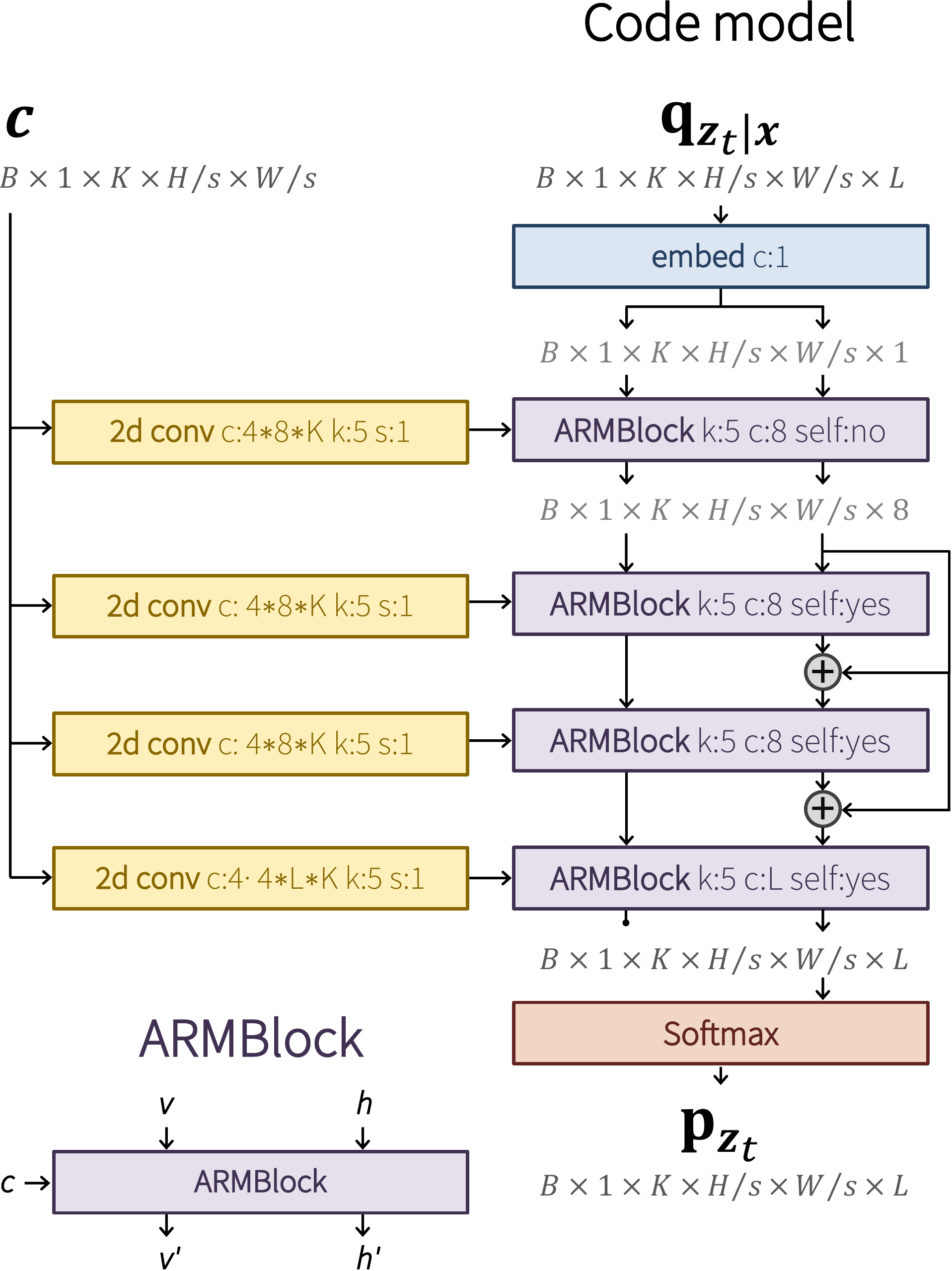}
    \vspace{5mm}
    \caption{Architecture of our prior / code-model. ARMBlock refers to a block in PixelCNN  \cite{vandenoordConditionalImageGeneration2016a}  with horizontal stack $h$, vertical stack $v$ and conditioning input $c$ (Figure 2 of \cite{vandenoordConditionalImageGeneration2016a}). $\mbf{c}$ represents the conditioning input that is used in our frame-conditioned and GRU-conditioned code-model.
    }
    \label{fig:model_architecture_prior}    
\end{figure}

We use a (conventional) convolutional layer to pre-process the conditioning input and to upsample the number of channels to match the size of each autoregressive block in the PixelCNN. The prior architecture is depicted in Figure \ref{fig:model_architecture_prior}. 


\subsubsection{Encoder and Code-Model gradients}
During training, the code model is updated to minimize the rate loss $\mc{L}_\text{rate} = \CE [q(\z | \x), p(\z) ]$. The rate loss is a sum over the elementwise cross-entropy between $q$ and $p$ (which is summed over each class $i$ of the codebook and each element $j$ of $\z$).

\begin{align}
\mc{L}_\text{rate}
&=
- \sum_\z q(\z | \x)  \log p(\z)
\\
&=
 - \sum_{ij} q_{\z|\x}^{ij}  \log p^{ij}_\z(\q_{\z|\x})
 = 
 \sum_{ij} \mc{L}_\text{rate}^{ij}
  \end{align}
  
Note that unlike \cite{mentzerConditionalProbabilityModels2018} we do not do any detaching of the gradient. As a result, the derivative of the rate loss w.r.t. the encoder parameters $\theta_q$ is affected by the code-model in the following way:

\begin{align}
 \pder{  \mc{L}_\text{rate}^{ij} }   {\theta_q}
     &= 
 -  \pder{  q_{\z|\x}^{ij} }   {\theta_q}  \left (   \dfrac{q_{\z|\x}^{ij} }{ p^{ij}_\z(\q_{\z|\x})}   \pder{ p^{ij}_\z(\q_{\z|\x})  }{q_{\z|\x}^{ij}}  
  +  \log p^{ij}_\z(\q_{\z|\x})  \right )
 \end{align}

Thus, the encoder is trained not only to minimize the distortion, but also to minimize the rate (e.g. to predict latents that are easily predictable by the code-model). This can also been seen by reversing the paths of the forward arrows in Figure \ref{fig:model-training}.

\section{Evaluation procedure}
\subsection{Traditional codec baselines}
We use FFMPEG\footnote{\url{https://ffmpeg.org/}}  2.8.15-0 to obtain the performance for the H.264/AVC and H.265/HEVC baselines. We use the default settings unless reported otherwise.

\subsection{Data preprocessing}
We build our evaluation datasets by extracting the png frames from the raw source videos using FFMPEG. Because some videos are in yuv colorspace, the conversion to rgb could in theory lead to some distortion, though this is imperceptible in practice. We use the same dataloading pipeline to evaluate our neural networks and the FFMPEG baselines as to avoid differences in ground-truth data.

\begin{figure*}[h!]
  \centering
  \begin{minipage}[t]{0.46\textwidth}
    \centering
    \includegraphics[width=\textwidth]{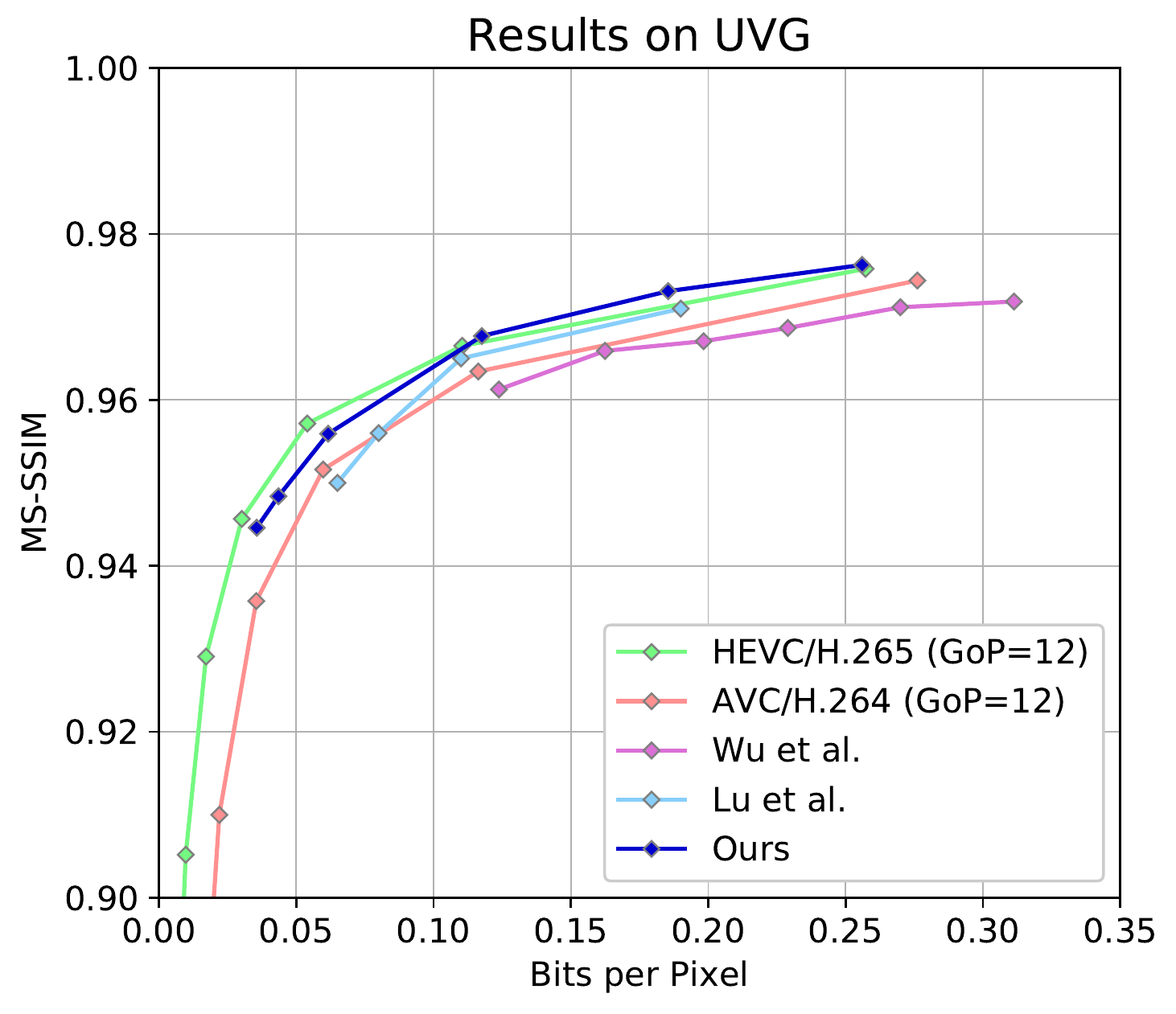}
    \caption{Rate/distortion results on UVG for classical and learned compression methods. H.264 and H.265 results were obtained with restricted FFMPEG settings (Group of Pictures set to 12).}
    \label{fig:standard_compression_uvg_restricted}
  \end{minipage}
  \hfill
 \begin{minipage}[t]{0.46\textwidth}
    \centering
    \includegraphics[width=\textwidth]{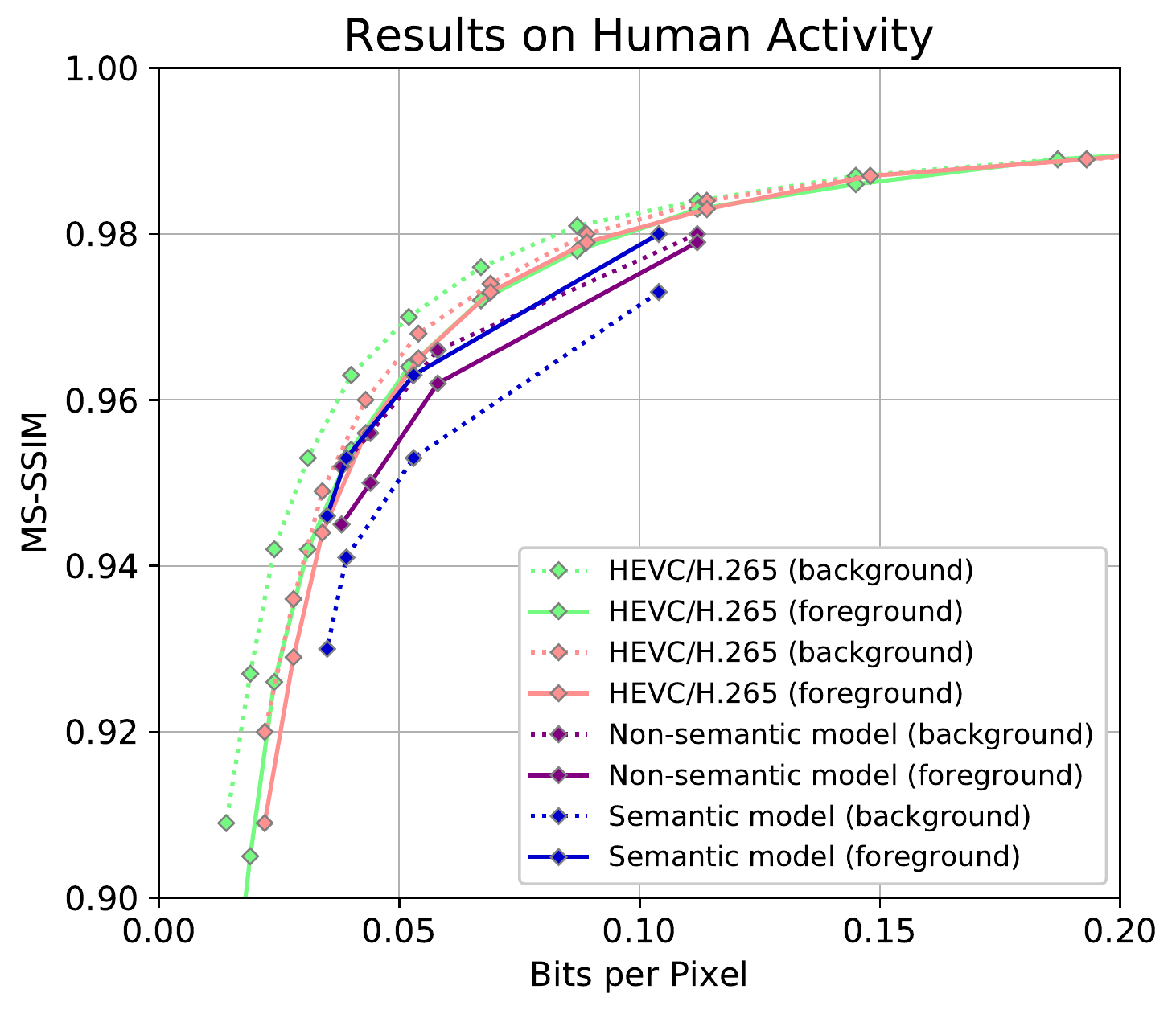}
    \caption{Semantic compression. Background is easier for all models, HEVC, AVC, and non-semantic model, except for our learned semantic compresison.}
    \label{fig:semantic_compression_full}
  \end{minipage}
\end{figure*}

\subsection{Rate-Distortion}
For the FFMPEG baselines, we divide the total filesize by the total number of pixels to obtain bpp. For our neural network, we use the rate loss (converted into bpp) as a proxy for rate. By definition, the rate loss gives the expected bitrate under adaptive arithmetic coding, and expected bpp was shown to be highly correlated to actual bpp \cite{luDVCEndtoendDeep2018}.

We calculate MS-SSIM \cite{wangMultiScaleStructuralSimilarity2003} using our own implementation which we benchmarked against the implementation in tensorflow\footnote{\url{https://www.tensorflow.org/versions/r1.13/api_docs/python/tf/image/ssim_multiscale}}. We use the same power factors that are initially proposed in \cite{wangMultiScaleStructuralSimilarity2003}.

\section{Additional results}

\subsection{Comparison to other methods}
When comparing neural networks to traditional codecs, it is common practice to evaluate those codecs under restrictive settings. For example, group of pictures (GoP) is often set to a value that is similar to the number of frames used to evaluate the neural networks \cite{wuVideoCompressionImage2018, luDVCEndtoendDeep2018}. Furthermore, encoding \texttt{preset} will be set to  \texttt{fast} (which will result in worse compression performance) \cite{wuVideoCompressionImage2018, luDVCEndtoendDeep2018}. In our evaluation (presented in Figure \ref{fig:standard_compression_uvg}) we instead use the FFMPEG default values of \texttt{GoP=25} and \texttt{preset=medium}. 

In Figure \ref{fig:standard_compression_uvg_restricted} we compare our end-to-end method to other learned compression methods and use baseline codecs with the restrictive setting of \texttt{GoP=12}, which is used in \cite{wuVideoCompressionImage2018, luDVCEndtoendDeep2018}. The figure shows that our model has a better rate-distortion performance than H.265/HEVC under these restrictive settings for bitrates higher than 1.2 bpp.

\subsection{Semantic Compression}
The results for semantic compression are reported in Figure \ref{fig:semantic_compression}. To avoid clutter, background performance for H.264/AVC and H.265/HEVC are omitted there. Figure \ref{fig:semantic_compression_full} shows the full results including background performance for traditional codecs.

\subsection{Adaptive Compression}
\begin{figure*}
    \centering
    \subfloat[HEVC/H.265  (0.025 BPP)] {\includegraphics[scale=0.5]{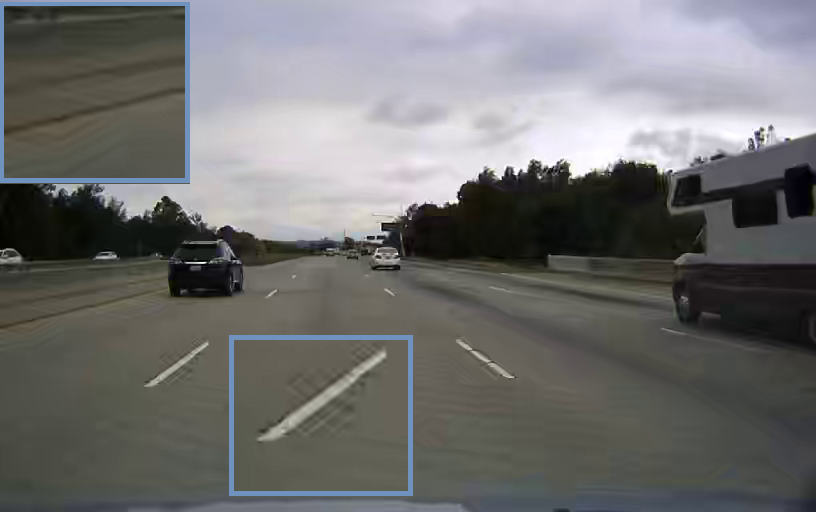}}\\
    \subfloat[Generic model (0.030 BPP)] {\includegraphics[scale=0.5]{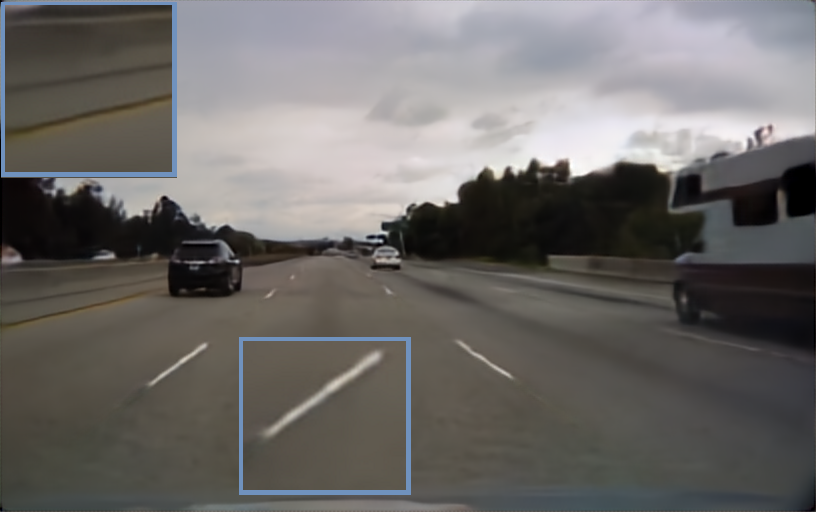}}\\
    \subfloat[Adapted model (0.025 BPP)] {\includegraphics[scale=0.5]{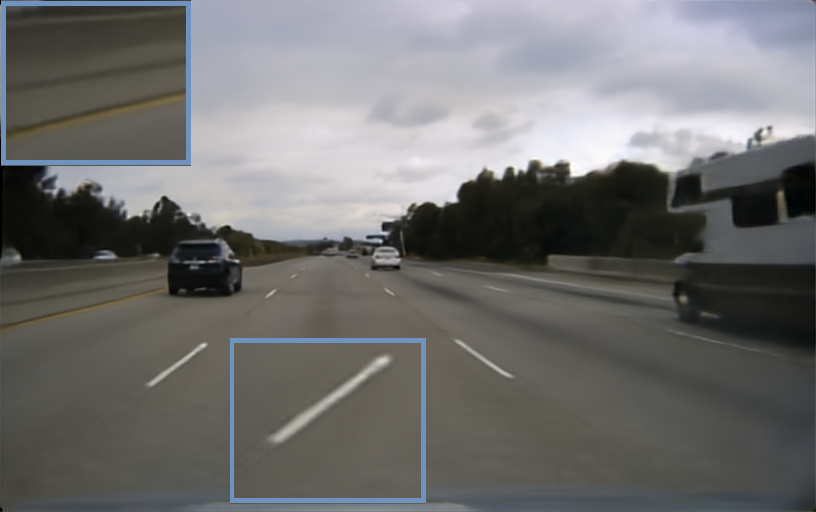}}
    \caption{Qualitative results for adaptive compression.}
    \label{fig:qualitative_adaptive}
\end{figure*}

Quantitative rate-distortion performance for adaptive compression is reported in Figure \ref{fig:adaptive_compression}. In Figure \ref{fig:qualitative_adaptive} we show a qualitative sample. Notice the clear block artifacts that can be observed around the road markings for H.265/HEVC. For our generic model, we do not observe such articacts, though we can see that edges are somewhat blurry around the line markings. In our adapted model, the road markings are significantly improved.

We note that the expanding perspective motion observed in road-driving footage is a great example of a predictable pattern in the data that a neural network could learn to exploit, while it would be difficult to manually engineer algorithms that use these patterns.

